\documentclass{rsproca}

\usepackage{graphicx}
\usepackage{amsmath}
\usepackage{amsfonts}
\usepackage{bm}
\usepackage{bbm}
\usepackage{url}
\usepackage{amssymb}
\usepackage{mathrsfs}
\usepackage{dsfont}
\usepackage{subfigure}
\usepackage{paralist}

\usepackage{stmaryrd}

\newcommand{\zero}{\bm{0}}
\newcommand{\vel}{\bm{v}}
\newcommand{\ba}{\bm{a}}

\newcommand{\ut}{\bm{t}}
\newcommand{\tp}{\bm{t}^+}
\newcommand{\tm}{\bm{t}^-}

\newcommand{\x}{\bm{x}}

\newcommand{\ex}{\bm{e}_x}
\newcommand{\ey}{\bm{e}_y}
\newcommand{\jump}[1]{\mbox{$\llbracket #1\rrbracket$}}

\newcommand{\Ws}{W_\mathrm{s}}
\newcommand{\Wstarp}{W^\ast_+}
\newcommand{\Wstarm}{W^\ast_-}
\newcommand{\Wstarpm}{W^\ast_\pm}

\newcommand{\taut}{\tau\ut}

\newcommand{\sspeed}{\dot{s}_0}
\newcommand{\sspspeed}{\dot{s}^\ast_+}
\newcommand{\ssmspeed}{\dot{s}^\ast_-}
\newcommand{\sspmspeed}{\dot{s}^\ast_\pm}
\newcommand{\sstarp}{s^\ast_+}
\newcommand{\sstarm}{s^\ast_-}

\newcommand{\force}{\bm{f}}

\newcommand{\Force}{\bm{F}_0}

\newcommand{\taup}{\tau^+}
\newcommand{\taustarp}{\tau^\ast_+}
\newcommand{\taustarpm}{\tau^\ast_\pm}
\newcommand{\Phistarp}{\bm{\Phi}^\ast_+}
\newcommand{\Phistarm}{\bm{\Phi}^\ast_-}
\newcommand{\Phistarpm}{\bm{\Phi}^\ast_\pm}
\newcommand{\tstarp}{\bm{t}^\ast_+}
\newcommand{\tstarm}{\bm{t}^\ast_-}
\newcommand{\tstarpm}{\bm{t}^\ast_\pm}
\newcommand{\taum}{\tau^-}
\newcommand{\taustarm}{\tau^\ast_-}
\newcommand{\dis}{\mathscr{D}}

\newcommand{\dray}{\delta\ray}
\newcommand{\ray}{\mathscr{R}}
\newcommand{\work}{\mathscr{W}}
\newcommand{\kin}{\mathscr{K}}
\newcommand{\dwork}{\delta\work}
\newcommand{\worka}{\mathscr{W}^{(\mathrm{a})}}
\newcommand{\worke}{\mathscr{W}^{(\mathrm{e})}}
\newcommand{\workc}{\mathscr{W}^{(\mathrm{c})}}
\newcommand{\dvel}{\delta\vel}
\newcommand{\dspeed}{\delta\sspeed}
\newcommand{\eucl}{\mathcal{E}}

\newcommand{\pt}{\partial_t}
\newcommand{\ps}{\partial_s}
\newcommand{\vnot}{\vel_0}
\newcommand{\xnot}{\x_0}
\newcommand{\xstarp}{\x^\ast_+}
\newcommand{\xstarm}{\x^\ast_-}
\newcommand{\vp}{\vel^+}
\newcommand{\vm}{\vel^-}
\newcommand{\vstarp}{\vel^\ast_+}
\newcommand{\vstarm}{\vel^\ast_-}

\newcommand{\ints}{\int_{s_1}^{s_2}}

\newcommand{\Wtwo}{\left(\tau\ut\cdot\vel\right)_{s=s_2}}
\newcommand{\Wone}{\left(\tau\ut\cdot\vel\right)_{s=s_1}}
\newcommand{\dvone}{\left(\dvel\right)_{s=s_1}}
\newcommand{\dvtwo}{\left(\dvel\right)_{s=s_2}}
\newcommand{\tautone}{\left(\taut\right)_{s=s_1}}
\newcommand{\tauttwo}{\left(\taut\right)_{s=s_2}}
\newcommand{\tautvone}{\left(\taut\cdot\vel\right)_{s=s_1}}

\newcommand{\jvel}{\jump{\vel}}
\newcommand{\jt}{\jump{\ut}}
\newcommand{\jtau}{\jump{\tau}}
\newcommand{\sgn}{\mathrm{sgn}}
\newcommand{\ydot}{\dot{y}}
\newcommand{\yddot}{\ddot{y}}
\newcommand{\tf}{t_\mathrm{f}}
\newcommand{\Pf}{P_\mathrm{f}}

\begin{document}

\title{Chain Paradoxes}

\author{
Epifanio G. Virga}

\address{Dipartimento di Matematica, Universit\`a di Pavia, Via Ferrata 5, I-27100 Pavia, Italy}

\subject{46.05.+b; 46.70.Hg}

\keywords{One-dimensional continua; Singularities; Shock waves; Dissipative dynamics; Chain dynamics.}

\corres{EG Virga\\
\email{eg.virga@unipv.it}}

\begin{abstract}
For nearly two centuries the dynamics of chains have offered examples of paradoxical theoretical predictions. Here we propose a theory for the dissipative dynamics of one-dimensional continua with singularities which provides a unified treatment for chain problems that have suffered from paradoxical solutions. These problems are duly solved within the present theory and their paradoxes removed---we hope.
\end{abstract}


\begin{fmtext}
\section{Introduction}\label{sec:intro}
There are two meanings ordinarily associated with the noun \emph{paradox}. Either it is an apparently absurd or counter-intuitive statement that further explanation nevertheless proves to be well-founded, or it is a statement that is taken to be self-contradictory or intrinsically unreasonable (adapted from \cite[p.\,185]{oed}). Often, in the former meaning, a paradox is also said to be \emph{apparent}. In brief, the intent of this paper is is to show that a number of paradoxes in chain dynamics are indeed apparent paradoxes, as they can be explained within a unified theory.

To this end, in Sections~\ref{sec:dissipation} and \ref{sec:shocks}, building on earlier work of O'Reilly and Varadi~\cite{reilly:treatment} and others before them, we propose a theory for dissipative singularities in a one-dimensional inextensible continuum, which we conventionally call a \emph{string}. The chain problems to which this theory is to be applied are presented in Section~\ref{sec:chain_problems} along with a brief history of the (supposedly) paradoxical solutions that have been proposed so far. In due order, these very problems are then solved within the present theory in Sections~\ref{sec:falling}, \ref{sec:folded}, and \ref{sec:sliding}. Although attained in a different perspective, the conclusions reached here are in tune with those of Grewal, Johnson, and Ruina~\cite{grewal:chain}, and of Hamm and G\'eminard~\cite{hamm:weight} before them. There is a common ground between these approaches which perhaps deserves further study and experimental validation.
\end{fmtext}
\maketitle

\section{Dissipation Principle for Singular Strings}\label{sec:dissipation}
Here we derive the balance equations that govern the dynamic evolution of inextensible strings with singularities in the velocity field from an appropriate extension of Rayleigh's dissipation principle.
The development illustrated below largely rests on Sect.~2.2 of \cite{sonnet:dissipative}, whose major essential features will be briefly recalled whenever the need for self-consistency is more urgently felt. Rayleigh's dissipation principle finds its complete formulation within thermodynamics, where it is also known as the principle of \emph{minimum entropy production}, see \cite[Sect.~2.2.2]{sonnet:dissipative}. Here we shall only need its purely mechanical variant, which can be easily obtained by specializing (and partly extending) the treatment in Sect.~2.2.3 of \cite{sonnet:dissipative} to the purely isothermal case. A similar development suited for classical mechanical systems with a finite number of degrees of freedom has recently been proposed in \cite{virga:comment} in response to a different attitude towards non-conservative systems \cite{galley:classical}. Here we shall extend the approach suggested in \cite{virga:comment} to one-dimensional continua with singularities, showing that Rayleigh's formalism is indeed more far-reaching than commonly believed.

For a material body, we denote by $\worka$ the power expended by all active agencies directly applied to it from the exterior (and which are given explicitly in terms of the body's motion). Letting $\kin$ denote the kinetic energy (in a given inertial frame), we call
\begin{equation}\label{eq:net_working_definition}
\worke:=\worka-\dot{\kin}
\end{equation}
the \emph{net working}, as it represents the amount of external power still available to the body once that going into the motion has been accounted for. In general, $\worke$ can be expressed as a \emph{linear} functional in the \emph{generalized} velocities of the body (including the classical velocity field $\vel$ and any extra field describing the evolution of additional degrees of freedom). By duality, the specific form of $\worke$ also identifies the \emph{generalized forces} corresponding to the generalized velocities. In the presence of internal constraints, the \emph{total working} $\work$ is defined as
\begin{equation}\label{eq:total_working_definition}
\work:=\worke+\workc,
\end{equation}
where $\workc$ is the power expended by the (generalized) reactions enforced by the constraints.

The version of Rayleigh's principle that we posit here states that the true evolution of a body is identified by requiring that a functional $\ray$, representing a \emph{dissipation potential}, be such that $\ray-\work$ is stationary with respect to all virtual motions for which the generalized forces are held fixed.\footnote{In its classical formulation, this is a \emph{minimum} principle, but it need not be so here. We shall see below when it does.} Formally, this amounts to requiring that
\begin{equation}\label{eq:delta_equality_principle}
\dray=\dwork,
\end{equation}
provided that $\dwork$ is linear in the (generalized) velocity variations. Classically, $\ray$ is a \emph{quadratic}, positive semi-definite functional, related to the \emph{dissipation} $\dis$ in the body through
\begin{equation}\label{eq:classical_dissipation}
\dis:=\frac12\ray.
\end{equation}
Were this always the case, Rayleigh's principle would be a genuine \emph{minimum} principle, and $\dis$ is actually minimized by the true motion compared to all virtual ones subject to the same generalized forces.\footnote{The whole Chapt.~2 of \cite{sonnet:dissipative} is devoted to alternative formulations of the classical Rayleigh's principle in different contexts.} Here, for the class of problems at hand, we shall afford a generalization of \eqref{eq:classical_dissipation}.

Consider an inextensible string moving in the three-dimensional space $\eucl$. We denote by $(s,t)\mapsto\x(s,t)\in\eucl$ the \emph{motion} of the string relative to a reference configuration, described by the convected co-ordinate $0\leqq s\leqq\ell$, which by the constraint of inextensibility can be taken to represent the arc-length in both the reference configuration and the present one, the latter being the image $\x([0,\ell],t)$ at time $t$ of the former. The velocity field $\vel$ is defined as
\begin{equation}\label{eq:velocity_definition}
\vel(s,t):=\pt\x(s,t).
\end{equation}
Likewise,
\begin{equation}\label{eq:acceleration_definition}
\ba(s,t):=\pt\vel(s,t)
\end{equation}
is the acceleration field. By requiring that the length of any arc $s_1\leqq s\leqq s_2$ be preserved in time, we easily obtain that $\vel$ must obey the constraint
\begin{equation}\label{eq:inextensibility_constraint}
\ps\vel(s,t)\cdot\ut=0,
\end{equation}
where
\begin{equation}\label{eq:unit_vector}
\ut(s,t):=\ps\x(s,t)
\end{equation}
denotes the unit tangent field in the present configuration.

\begin{figure}[!h]
  \centering
  \includegraphics[width=.4\linewidth]{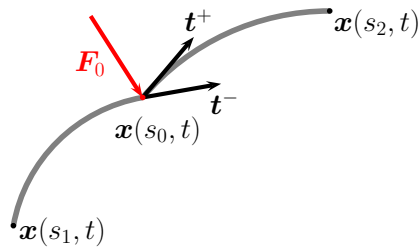}
  \caption{
A singular point at $\x_0(t)=\x_0(s_0(t),t)$, where the unit tangent $\ut$ jumps from $\tm$ to $\tp$ as the arc-length parameter $s$ traverses $s_0$, is acted upon by the force $\bm{F}_0$. The arc under consideration is described by the interval $s_1\leqq s\leqq s_2$.}
\label{fig:shock}
\end{figure}

We further assume that a \emph{shock} traverses the reference configuration, represented by a propagating front $s_0=s_0(t)$ upon which the velocity field $\vel$ is discontinuous, while $\x$ itself is continuous. The point $\x_0(t):=\x(s_0(t),t)$ designates the position in space of the shock. We call it a \emph{singular point}, for brevity; the arcs of the string that do not contain singular points will be called \emph{regular}. Letting for any function $\Psi$ denote by $\Psi^+$ and $\Psi^-$ the right and left limits of $\Psi$, that is, the limits for $s\to s_0^+$ and $s\to s_0^-$, respectively, we set, as usual,
\begin{equation}\label{eq:jump_definition}
\jump{\Psi}:=\Psi^+-\Psi^-.
\end{equation}
We denote by $\vnot$ the velocity of the singular point $\xnot$. It is not a material velocity and it must be continuous across the shock:
\begin{equation}\label{eq:v_0_expressions}
\vnot=\vp+\sspeed\tp=\vm+\sspeed\tm.
\end{equation}
Hence
\begin{equation}\label{eq:jump_condition_velocity}
\jump{\vnot}=\jump{\vel}+\sspeed\jump{\ut}=\zero.
\end{equation}
Figure \ref{fig:shock} illustrates the general situation envisaged here.
Equations \eqref{eq:inextensibility_constraint} and \eqref{eq:jump_condition_velocity} represent constraints that must be enforced at all times along the string and at the singular point $\xnot$, respectively.

To free the velocity field $\vel$ from its constraint we account for the power $\workc$ expended by the associated reactive force. We write $\workc$ in the form
\begin{equation}\label{eq:working_constraints}
\workc=-\ints\tau\ps(\vel\cdot\ut)ds,
\end{equation}
where $s_1$ and $s_2\geqq s_1$ delimit an interval which comprises the propagating front, $s_1\leqq s_0\leqq s_2$, and $\tau$ represents an unknown \emph{tension} along the string, which may suffer a jump at the singular point $\xnot$ and will eventually be determined by requiring \eqref{eq:inextensibility_constraint} to be satisfied. In a different fashion, we shall enforce \eqref{eq:jump_condition_velocity} by requiring $\vnot$ as defined by \eqref{eq:v_0_expressions} to be continuous across $\xnot$.
An integration by parts performed in both integrals obtained by splitting the integral in \eqref{eq:working_constraints} over the intervals $[s_1,s_0]$ and $[s_0,s_2]$ leads us to the following equivalent expression for $\workc$,
\begin{equation}\label{eq:working_constraints_equivalent}
\workc=\ints\ps(\taut)\cdot\vel ds+\jump{\taut\cdot\vel}-\Wtwo+\Wone.
\end{equation}

Letting $\force$ denote the force acting per unit length of the string and $\lambda$ its (uniform) linear mass density, if an external concentrated force $\Force$ is applied to $\xnot$ (as for the interaction with an obstacle), the net working $\worke$ can be written as
\begin{align}\label{eq:net_working_formula}
\worke=&\ints\force\cdot\vel ds+\Force\cdot\vnot-\frac{d}{dt}\ints\frac12\lambda v^2ds\nonumber\\
=&\ints\left(\force-\lambda\ba\right)\cdot\vel ds+\Force\cdot\vnot+\frac12\lambda\sspeed\jump{v^2},
\end{align}
which is easily derived by splitting as above the integral of the kinetic energy over the intervals $[s_1,s_0]$ and $[s_0,s_2]$.

For simplicity, we shall only consider a portion of string with a single singular point $\xnot$, which is also identified with the only source of dissipation.  The Rayleigh potential $\ray$, which is to be related to the dissipation $\dis$ in the body not necessarily through \eqref{eq:classical_dissipation}, reduces to a function of objective\footnote{That is, invariant under a change of observer, represented as a generic, time dependent orthogonal transformation.} dissipation measures. We take these latter to be the shock's speed $\sspeed$ in the reference configuration and the jump $\jump{\vel}$ of the the velocity across the singular point $\xnot$, $\ray=\ray(\sspeed,\jvel)$.

Two identities, which follow from \eqref{eq:jump_condition_velocity}, are instrumental to expressing the jump terms in both \eqref{eq:working_constraints_equivalent} and \eqref{eq:net_working_formula} in a way that makes only $\vnot$ appear there as the natural velocity against which power is expended at the singular point $\xnot$. These are as follows
\begin{subequations}\label{eq:v_0_identities}
\begin{equation}\label{eq:v_0_identity_a}
\jump{\taut\cdot\vel}=\jump{\taut}\cdot\vnot-\sspeed\jump{\tau},
\end{equation}
\begin{equation}\label{eq:v_0_identity_b}
\frac12\lambda\sspeed\jump{v^2}=-\lambda\sspeed^2\jump{\ut}\cdot\vnot,
\end{equation}
and are supposed to provide linear forms in $\vnot$ and $\sspeed$, so as to identify by duality the appropriate generalized forces expending power against these kinematic measures in \eqref{eq:working_constraints_equivalent} and \eqref{eq:net_working_formula}. We heed for later use that \eqref{eq:jump_condition_velocity} also allows one to rewrite \eqref{eq:v_0_identity_b} as
\begin{equation}\label{eq:v_0_identity_c}
\frac12\lambda\sspeed\jump{v^2}=\lambda\sspeed\jvel\cdot\vnot.
\end{equation}
\end{subequations}

By using both \eqref{eq:v_0_identity_a} and \eqref{eq:v_0_identity_b} in \eqref{eq:working_constraints_equivalent} and \eqref{eq:net_working_formula}, and performing arbitrary variations $\dspeed$ and $\dvel$ of both $\sspeed$ and $\vel$, respectively, while leaving all generalized forces frozen, we arrive at
\begin{equation}\label{eq:total_working_variation}
\dwork=\Force\cdot\dvel_0-\lambda\sspeed^2\jump{\ut}\cdot\dvel_0+\jump{\taut}\cdot\dvel_0-\jump{\tau}\dspeed
+\ints\left[\force-\lambda\ba+\ps(\taut)\right]\cdot\dvel ds,
\end{equation}
where use has also been made of the conditions
\begin{equation}\label{eq:delta_v_1_2_vanish}
\dvone=\dvtwo=\zero,
\end{equation}
which limit the variation of the motion to the portion of string under consideration. Correspondingly,
\begin{equation}\label{eq:rayleigh_variation}
\dray=\frac{\partial\ray}{\partial\sspeed}\dspeed+\frac{\partial\ray}{\partial\jvel}\cdot\jump{\dvel},
\end{equation}
which follows from remarking that $\delta\jvel=\jump{\dvel}$ and by \eqref{eq:jump_condition_velocity} is further changed into
\begin{equation}\label{eq:delta_Rayleigh}
\delta\ray=\left(\frac{\partial\ray}{\partial\sspeed} -\frac{\partial\ray}{\partial\jvel}\cdot\jump{\ut}\right)\dspeed.
\end{equation}
Requiring the variational principle in \eqref{eq:delta_equality_principle} to be identically valid for arbitrary variations $\dvel$ and $\dspeed$, we arrive at the equations
\begin{subequations}\label{eq:evolution_equations}
\begin{equation}\label{eq:balance_equation_linear_momentum}
\lambda\ba=\force+\ps(\taut)\quad\text{in}\quad [s_1,s_2],
\end{equation}
\begin{equation}\label{eq:evolution_equation_v_0}
\jump{\taut}+\Force+\lambda\sspeed\jvel=\zero,
\end{equation}
\begin{equation}\label{eq:evolution_equation_s_0}
-\jtau=\frac{\partial\ray}{\partial\sspeed}-\frac{\partial\ray}{\partial\jvel}\cdot\jt,
\end{equation}
\end{subequations}
where \eqref{eq:v_0_identity_c} has also been used. These are the evolution equations of the theory; they are valid along any regular arc and at any singular point of an inextensible string.
While \eqref{eq:balance_equation_linear_momentum} is clearly the expected balance equation for linear momentum on regular arcs of the string, with $\tau$ the internal tension, \eqref{eq:evolution_equation_v_0} and \eqref{eq:evolution_equation_s_0} are still to be related to the classical balances of linear momentum and energy at a singular point. We shall see shortly below how this can be achieved. I wish first employ equations \eqref{eq:evolution_equations} to relate the Rayleigh potential $\ray$ to the dissipation $\dis$ in the system.

By use of \eqref{eq:balance_equation_linear_momentum} and \eqref{eq:evolution_equation_v_0} in \eqref{eq:working_constraints_equivalent} and \eqref{eq:net_working_formula}, with the aid of the identities \eqref{eq:v_0_identity_a} and \eqref{eq:v_0_identity_c}, we can rewrite $\work$ as
\begin{equation}\label{eq:total_working_rewritten}
\work+\Wtwo-\Wone=-\sspeed\jtau,
\end{equation}
which by \eqref{eq:evolution_equation_s_0} and \eqref{eq:jump_condition_velocity} becomes
\begin{equation}\label{eq:total_working_rewritten_bis}
\work+\Wtwo-\Wone=\dis,
\end{equation}
where
\begin{equation}\label{eq:dissipation_definition}
\dis:=\frac{\partial\ray}{\partial\sspeed}\sspeed+\frac{\partial\ray}{\partial\jvel}\cdot\jvel
\end{equation}
clearly acquires the meaning of the energy dissipated per unit time in the string, since the singular point $\xnot$ is admittedly the only point where energy may get lost.

By its very physical interpretation, we shall consider $\dis$ as the primary constitutive function of the theory; the dissipation potential $\ray$, which enters the evolution equation \eqref{eq:evolution_equation_s_0} is to be derived from $\dis$ through \eqref{eq:dissipation_definition}. Solving this equation by the method of characteristics, we arrive at the following explicit expression for $\ray$, to within an arbitrary additive constant,
\begin{equation}\label{eq:rayleigh_explicit}
\ray(\sspeed,\jvel)=\left.\int\dis\left(e^\sigma\sspeed,e^\sigma\jvel\right)d\sigma\right|_{\sigma=0}.
\end{equation}
By comparing \eqref{eq:total_working_rewritten} and \eqref{eq:total_working_rewritten_bis}, we easily see that whenever $\sspeed\neq0$ equation \eqref{eq:evolution_equation_s_0} can equivalently be written as
\begin{equation}\label{eq:dissipation_gold}
\sspeed\jtau=-\dis,
\end{equation}
which has the advantage of linking directly the jump in tension to the dissipated energy.

It follows from \eqref{eq:rayleigh_explicit} that whenever $\dis$ is homogeneous of degree $n$ in all its variables,
\begin{equation}\label{eq:rayleigh_n_dissipation}
\ray=n\dis,
\end{equation}
of which \eqref{eq:classical_dissipation} is just a special case. In the present context, there is no guarantee that \eqref{eq:classical_dissipation} holds. More generally, we shall assume that $\dis$ is positive semi-definite and vanishing when \emph{either} $\sspeed$ \emph{or} $\jvel$ vanish. For such a $\dis$, $\ray$ as delivered by \eqref{eq:rayleigh_explicit} identifies through \eqref{eq:evolution_equation_s_0} the evolution equation for $s_0$ required by the theory. This equation together with \eqref{eq:jump_condition_velocity} and \eqref{eq:balance_equation_linear_momentum} will determine the motion of a singular string.

Equation \eqref{eq:rayleigh_n_dissipation} is not expected to be valid in general. Nonetheless, the  proposal for $\dis$ made in \cite{virga:dissipative}, whereas does not comply with \eqref{eq:classical_dissipation}, still makes \eqref{eq:rayleigh_n_dissipation} satisfied. It was assumed in \cite{virga:dissipative} that
\begin{equation}\label{eq:dissipation_example_1}
\dis=\frac12f\lambda|\sspeed|\jvel^2,
\end{equation}
with $0\leqq f\leqq1$ a phenomenological, dimensionless coefficient. It readily follows from \eqref{eq:rayleigh_explicit} that the Rayleigh function $\ray$ associated with this $\dis$ is given by \eqref{eq:rayleigh_n_dissipation} with $n=3$. The expression for $\dis$ in \eqref{eq:dissipation_example_1} will also be adopted in the rest of the paper, for it is amenable to a physical interpretation based on the classical laws of impact dynamics (see Section~4\ref{sec:dissipation_chains} below).

\subsection{Classical Balances}\label{sec:classical}
We close this section by showing how equation \eqref{eq:evolution_equation_v_0} and \eqref{eq:evolution_equation_s_0} are related to the classical balances of linear momentum and energy at the singular point $\xnot$. That \eqref{eq:evolution_equation_v_0} expresses precisely the former follows from writing as in \cite{reilly:treatment} the balance of linear momentum for the whole arc $[s_1,s_2]$ in the form
\begin{equation}\label{eq:balance_linear_momentum_whole_arc}
\pt\ints\lambda\vel ds=\tauttwo-\tautone+\ints\force ds+\Force.
\end{equation}
Splitting the integral on the left side of this equation over the intervals $[s_1,s_0]$ and $[s_0,s_2]$, both of which depend on time, and taking the limits as both $s_2\to s_0^+$ and $s_1\to s_0^-$, which isolates the singular point $\xnot$ from the regular arcs adjacent to it, we reobtain \eqref{eq:evolution_equation_v_0}.

In a similar way, granted that the dissipation $\dis$ is concentrated at $\xnot$, the balance of energy over the whole arc $[s_1,s_2]$ reads as
\begin{equation}\label{eq:balance_energy_whole_arc}
\pt\frac12\ints\lambda v^2ds+\dis=\Wtwo-\Wone+\ints\force\cdot\vel ds+\Force\cdot\vnot,
\end{equation}
whence by the same reasoning as above we arrive at the following localized energy balance,
\begin{equation}\label{eq:balance_energy_localized}
\jump{\taut\cdot\vel}+\Force\cdot\vnot+\frac12\lambda\sspeed\jump{v^2}=\dis.
\end{equation}
For $\Force=\zero$, this balance equation coincides with (5) of \cite{virga:dissipative}, provided that we identify the power supply $\Ws$ introduced there at $\xnot$ with $-\dis$. Likewise, it reduces to the analogous equation (2.7)$_3$ of \cite{reilly:treatment}, provided that there $\phi_E$  incorporates both the energy dissipated at the singular point and the power expended by the external force. Equation \eqref{eq:balance_energy_localized} also agrees with (13) of \cite{hanna:jump}, provided that the power supply $E$ introduced there is written as $E=\dis-\Force\cdot\vnot$. Now, \eqref{eq:balance_energy_localized} is nothing but \eqref{eq:dissipation_gold}, as shown by using both \eqref{eq:v_0_identity_a} and \eqref{eq:v_0_identity_c} and combining the result with \eqref{eq:evolution_equation_v_0}. This also shows that \eqref{eq:evolution_equation_v_0} and \eqref{eq:evolution_equation_s_0} together imply \eqref{eq:balance_energy_localized}. Thus, our interpretation of the evolution equations \eqref{eq:evolution_equation_v_0} and \eqref{eq:evolution_equation_s_0} as balance laws is completely justified. Often in the following we shall turn to \eqref{eq:dissipation_gold} as an equivalent formulation for \eqref{eq:evolution_equation_s_0} in the presence of a propagating shock, and so our theory will be based on \eqref{eq:balance_equation_linear_momentum}, \eqref{eq:evolution_equation_v_0}, and \eqref{eq:dissipation_gold}.

\section{External Shocks}\label{sec:shocks}
The shocks considered in the preceding section, which manifest themselves at singular points in the middle of a moving string, will be called \emph{internal} to distinguish them from the \emph{external} shocks considered here. An external shock manifests itself whenever a moving string gets in contact with a still, shapeless reservoir from where it is ejected and set in motion or where it comes to an abrupt halt. This will be our way of treating continuous systems with a variable mass. Essentially, this idea goes back to Cayley~\cite{cayley:class}, who introduced the notion of \emph{continuously} imparted impacts, equally able of setting resting matter in motion and of putting moving matter at rest. Though our derivation of the balance equations governing an external shock will be similar to the one presented in Section~2\ref{sec:classical} for internal shocks, the reader should not be deceived into thinking that an external shock is nothing but an internal shock with zero velocity on one side. In an external shock the string is supposed to be quiescent on one side of the shock, but its shape is completely undetermined and it only plays the role of providing sources for both linear momentum and energy.

\begin{figure}[!h]
  \centering
  \subfigure[]{\includegraphics[width=.25\linewidth]{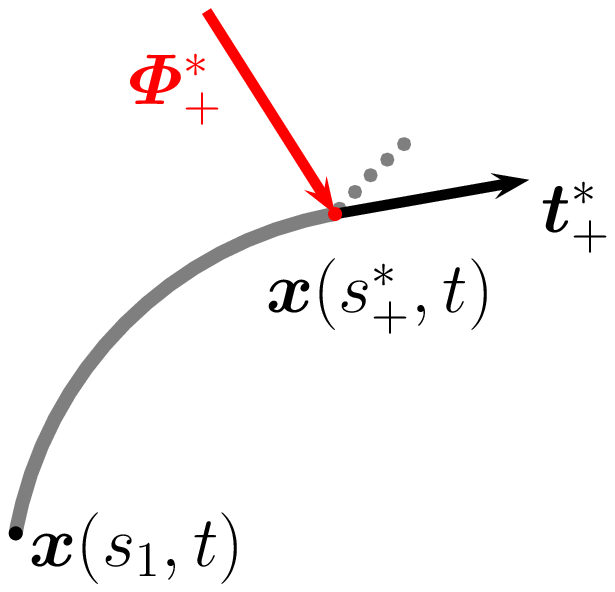}}
  \hspace{0.1\linewidth}
  \subfigure[]{\includegraphics[width=.4\linewidth]{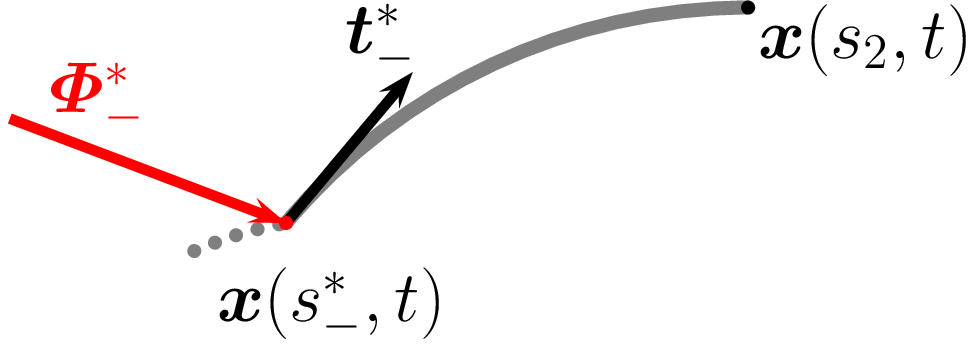}}
  \caption{External shocks adjacent to a quiescent string. (a) The quiescent string following the moving string is represented by incoherent dots to highlight the irrelevance of its shape to our development. The shock takes place at the singular point $\xstarp(t):=\x\left(\sstarp(t),t\right)$ and is sustained by a source, $\Phistarp$, of linear momentum; $\tstarp$ denotes the outer unit tangent of the moving string at the singular point. (b) The quiescent string precedes the moving string and is connected to it through the singular point $\xstarm(t):=\x\left(\sstarm(t),t\right)$, where the inward unit tangent to the moving string is denoted by $\tstarm$; $\Phistarm$ is the corresponding source of linear momentum.
}
\label{fig:external_shocks}
\end{figure}

We distinguish two types of external shocks, as to whether the quiescent string follows or precedes the moving string. Both situations are depicted in Fig.~\ref{fig:external_shocks}.
We denote by $\Phistarp$ and $\Wstarp$ the sources of linear momentum and energy, respectively, for the former case, and by $\Phistarm$ and $\Wstarm$ the corresponding sources for the latter case. We note in passing that $-\Phistarp$ and $-\Phistarm$ are the forces transferred to the quiescent string in the two case, while $-\Wstarp$ and $-\Wstarm$ are the corresponding energy sources for the quiescent string in the two cases.\footnote{Clearly, as we do not prescribe the signs of $\Wstarpm$, these may equally well represent energy gains or energy losses.} The external shock represented in Fig.~\ref{fig:external_shocks}(a) is characterized by a front $\sstarp(t)$ evolving in time in both the reference and present configurations, so that the string's velocity $\vel$ jumps from $\vstarp$ to nought at the singular point $\xstarp(t):=\x\left(\sstarp(t),t\right)$. By requiring the velocity $\vnot^+$ of the geometric point that instantaneously coincides with $\xstarp$ to vanish identically in time, as commanded by the kinematic compatibility with the quiescent string, we readily obtain that
\begin{equation}\label{eq:external_shock_kinematic_compatibility}
\vnot^+=\pt\x(\sstarp,t)=\vstarp+\sspspeed\tstarp=\zero,
\end{equation}
which here replaces \eqref{eq:v_0_expressions}.\footnote{Incidentally, \eqref{eq:external_shock_kinematic_compatibility} also ensures that the velocity $\vstarp$ must be tangent to the moving string, and so this must move along its length at least just before coming to an abrupt stop.}

The balance of linear momentum for the arc of string in Fig.~\ref{fig:external_shocks}(a), comprised between $s=s_1$ and $s=\sstarp$, can be written in the form
\begin{equation}\label{eq:external_shock_balance_linear_momentum}
\pt\int_{s_1}^{\sstarp}\lambda\vel ds=\Phistarp-\tautone+\int_{s_1}^{\sstarp}\force ds,
\end{equation}
where as in Section~\ref{sec:dissipation} $\force$ denotes the force acting per unit length of the string. Accounting in \eqref{eq:external_shock_balance_linear_momentum} for the dependence of $\sstarp$ on time and taking the limit as $s_1$ tends to $\sstarp$, we arrive at
\begin{equation}\label{eq:external_shock_pre_balance_linear_momentum_plus}
\lambda\sspspeed\vstarp=\Phistarp-\taustarp\tstarp,
\end{equation}
where $\taustarp$ is the tension in the moving string at the singular point $\xstarp$. By \eqref{eq:external_shock_kinematic_compatibility} we give \eqref{eq:external_shock_pre_balance_linear_momentum_plus} the following form
\begin{equation}\label{eq:external_shock_balance_linear_momentum_plus}
\Phistarp=\left[\taustarp-\lambda(\sspspeed)^2\right]\tstarp.
\end{equation}
If the balance of linear momentum in \eqref{eq:external_shock_balance_linear_momentum} echoes \eqref{eq:balance_linear_momentum_whole_arc}, the balance of energy in \eqref{eq:balance_energy_whole_arc} here becomes
\begin{equation}\label{eq:external_shock_balance_energy}
\pt\frac12\int_{s_1}^{\sstarp}\lambda v^2ds=-\tautvone+\int_{s_1}^{\sstarp}\force ds+\Wstarp,
\end{equation}
where account has also be taken of the fact that the force $\Phistarp$ expends no power since $\vnot^+=\zero$. By performing the time differentiation in \eqref{eq:external_shock_balance_energy} and then taking the limit as $s_1\to\sstarp$, also with the aid of \eqref{eq:external_shock_kinematic_compatibility}, we obtain the following balance law for the energy at the singular point $\xstarp$,
\begin{equation}\label{eq:external_shock_balance_energy_plus}
\Wstarp=-\left[\taustarp-\frac12\lambda(\sspspeed)^2\right]\sspspeed.
\end{equation}

Reasoning in exactly the same way, we derive the analogs of the kinematic condition \eqref{eq:external_shock_kinematic_compatibility} and balance equations \eqref{eq:external_shock_balance_linear_momentum_plus} and \eqref{eq:external_shock_balance_energy_plus} at the singular point $\xstarm$ depicted in Fig.~\ref{fig:external_shocks}(b) where a quiescent string is abruptly set in motion. Letting $\vstarm$ denote the velocity of the moving sting at $\xstarm$ and $\tstarm$ the inward unit tangent, we arrive at the following equations
\begin{subequations}
\begin{equation}\label{eq:external_shock_kinematic_compatibility_minus}
\vstarm+\ssmspeed\tstarm=\zero,
\end{equation}
\begin{equation}\label{eq:external_shock_balance_linear_momentum_minus}
\Phistarm=-\left[\taustarm-\lambda(\ssmspeed)^2\right]\tstarm,
\end{equation}
\begin{equation}\label{eq:external_shock_balance_energy_minus}
\Wstarm=\left[\taustarm-\frac12\lambda(\ssmspeed)^2\right]\ssmspeed,
\end{equation}
\end{subequations}
where $\taustarm$ is the tension in the moving string at $\xstarm$. In particular, kinematic compatibility requires the string to be injected at $\xstarm$ along its tangent and the balance of linear momentum requires the force $\Phistarm$ to be directed along the string's tangent as well.

Although equations \eqref{eq:external_shock_kinematic_compatibility_minus} and \eqref{eq:external_shock_balance_linear_momentum_plus} are valid in general, precisely as are \eqref{eq:external_shock_kinematic_compatibility} and \eqref{eq:external_shock_balance_linear_momentum_plus}, they all are compatible with a special class of motions for a string which will  be our object in the following sections, the one in which a string slides everywhere along its length. Such motions are characterized by the kinematic requirement
\begin{equation}\label{eq:motion_along_its_length}
\vel=v\ut,
\end{equation}
enforced along the entire moving string. It is compatible with the inextensibility constraint \eqref{eq:inextensibility_constraint} only is $v$ is assumed to be a function of time only, $v=v(t)$. Under this assumption, both \eqref{eq:external_shock_kinematic_compatibility} and \eqref{eq:external_shock_kinematic_compatibility_minus} reduce to
\begin{subequations}\label{eq:external_shock_plus_minus}
\begin{equation}\label{eq:external_shock_speeds}
\sspmspeed=-v
\end{equation}
and \eqref{eq:external_shock_balance_linear_momentum_plus}, \eqref{eq:external_shock_balance_energy_plus}, \eqref{eq:external_shock_balance_linear_momentum_minus}, and \eqref{eq:external_shock_balance_energy_minus} collectively simplify into
\begin{equation}\label{eq:external_shock_balance_linear_momentum_plus_minus}
\Phistarpm=\pm\left(\taustarpm-\lambda v^2\right)\tstarpm,
\end{equation}
\begin{equation}\label{eq:external_shock_balance_energy_plus_minus}
\Wstarpm=\pm\left(\taustarpm-\frac12\lambda v^2\right)v.
\end{equation}
\end{subequations}
These equations will often be called upon in the following.

\section{Chain Problems}\label{sec:chain_problems}
In the following sections we shall apply the general theory presented so far to three specific problems in the dynamics of chains. Below in this section the reader will be introduced to these problems collectively, also indulging in some historical notes on the supposedly paradoxical aspects that they have shown since the work of Buquoy nearly two centuries ago.\footnote{As recounted by \v{S}\'{\i}ma and Podolsk\'y~\cite{sima:buquoy}, in August 1815 Graf Georg von Buquoy presented his theory on the motion of systems with a variable mass at the Paris Academy of Sciences. Buquoy's problem is indeed similar to the problem of a falling chain treated in Section~\ref{sec:falling}. If attention is restricted to an appropriate subsystem, all the problems treated here could also be seen as problems for systems with a variable mass.} Before doing so, however, we need supplement the theory by showing that the choice for the shock dissipation $\dis$ in \eqref{eq:dissipation_example_1} is the most appropriate for chains.

\subsection{Dissipation in Chains}\label{sec:dissipation_chains}
Resuming the notation introduced in \cite{virga:dissipative}, for a shock in a chain we shall denote by $\Ws$ the energy supply at a singular point, and so hereafter we shall set
\begin{equation}\label{eq:W_s_formula_chain}
\Ws:=-\dis=-\frac12f\lambda|\sspeed|\jvel^2.
\end{equation}

The kink formed in a chain where a shock is propagating through it is likely to involve an impact between the chain's links that come there in contact. When two smooth rigid bodies collide in a single point of their boundaries, the variation $\Delta T$ in the total kinetic energy during the impact can be expressed as
\begin{equation}\label{eq:Delta_T_impact}
\Delta T=-\frac{1-e}{1+e}T_\Delta,
\end{equation}
where $e$ is Newton's restitution coefficient and $T_\Delta$ is the kinetic energy of the velocity differences, that is, the kinetic of the imaginary motion\footnote{Which in \cite{virga:dissipative} was also referred to as the \emph{lost motion}.} that would result from attributing to all the particles in the colliding bodies the velocities $\Delta\vel_i$ they lose in the impact. In other words, letting $\vel^-_i$ be the velocity of the $i$th particle before impact and $\vel^+_i$ that after impact, $\Delta\vel_i:=\vel^+_i-\vel^-_i$ and
\begin{equation}\label{eq:T_Delta_impact}
T_\Delta=\frac12\sum_im_i(\Delta\vel_i)^2,
\end{equation}
where $m_i$ is the mass of the $i$th particle and the sum is extended to any discretization of of both colliding bodies.\footnote{Equation \eqref{eq:Delta_T_impact} is proved directly in \cite[p.\,235]{whittaker:treatise} and also derived in \cite[p.\,743]{levi_civita:lezioni} from the classical theory of Poisson for the impact between smooth rigid bodies.}

When a chain is described as a one-dimensional, inextensible string, as we did above, and a shock propagating though it is envisaged as a localized mass transfer with rate $\lambda|\sspeed|$ associated with the energy supply $\Ws$, \eqref{eq:T_Delta_impact} directly suggests taking the rate of change of $T_\Delta$ as
\begin{equation}\label{eq:rate_T_Delta_impact}
\dot{T}_\Delta=\frac12\lambda|\sspeed|\jvel^2,
\end{equation}
so that \eqref{eq:Delta_T_impact} the justifies \eqref{eq:W_s_formula_chain}, after setting
\begin{equation}\label{eq:f_function_of_e}
f=\frac{1-e}{1+e}.
\end{equation}
In the following we shall not insist in representing the coefficient $f$ through \eqref{eq:f_function_of_e} in terms of the restitution coefficient for the impact of two chain's links. Rather, we shall consider $f$ as a phenomenological coefficient pertaining only to the shape of the links in a chain and the material that constitutes them. Of \eqref{eq:f_function_of_e} we shall only retain the bounds it imposes on $f$, $0\leqq f\leqq1$,\footnote{Assuming, as usually done, that $0\leqq e\leqq1$.} and we shall say that $f=0$ is the completely elastic limit, as there would be no dissipation at the shock, while $f=1$ is the completely inelastic limit, for which the dissipation at the shock would be maximum, for a given velocity jump $\jvel$. As we shall see, $f$ is a measurable parameter, and indeed it has been measured already, though perhaps inadvertently.

\subsection{The Problems}\label{sec:historical_notes}
How chains have come to be associated with paradoxes is quite an interesting story, which here I only recount briefly. These \emph{paradoxes} are indeed counter-intuitive predictions which some theories made and others did not support, which were sometimes sustained by experimental observations and sometimes were not. Often it was considered paradoxical in this context relaxing an assumption mostly taken for granted and regarded as natural, but indeed unnecessary. The examples treated in the following are concerned with chains falling in one fashion or another in the terrestrial gravitational field. For them, for example, it might have been regarded as paradoxical a theory that predicted falling with an acceleration greater than the gravitational acceleration $g$, not to mention the possibility that in the continuum limit either the velocity or the acceleration became unbounded during the motion.

\begin{figure}[!h]
  \centering
  \subfigure[]{\includegraphics[width=.3\linewidth]{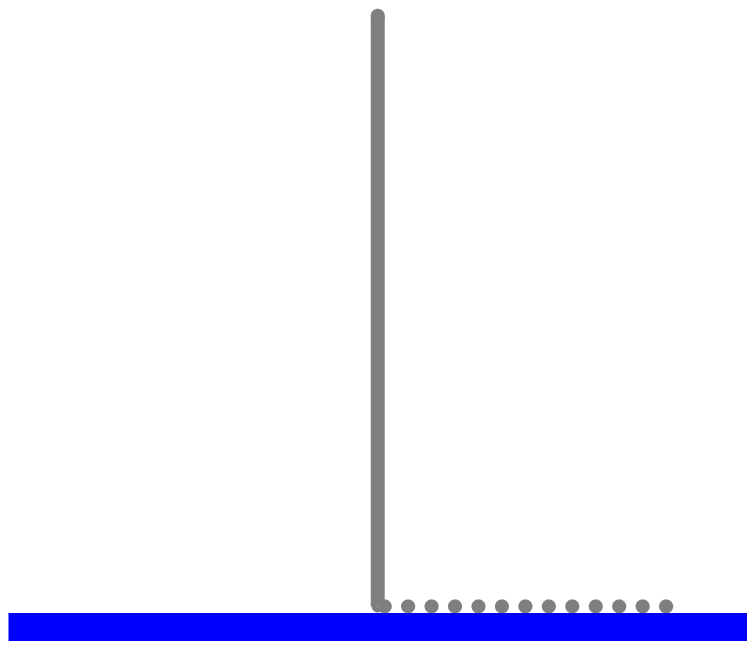}}
  \hspace{0.1\linewidth}
  \subfigure[]{\includegraphics[width=.15\linewidth]{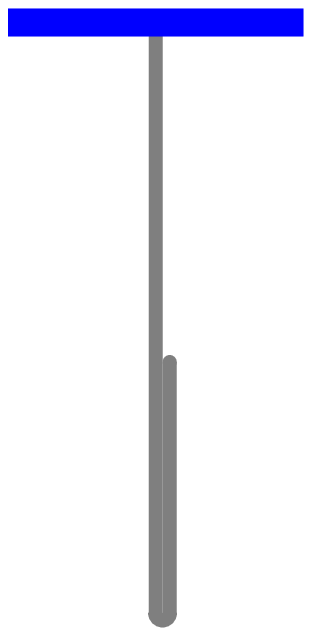}}
  \hspace{0.1\linewidth}
  \subfigure[]{\includegraphics[width=.3\linewidth]{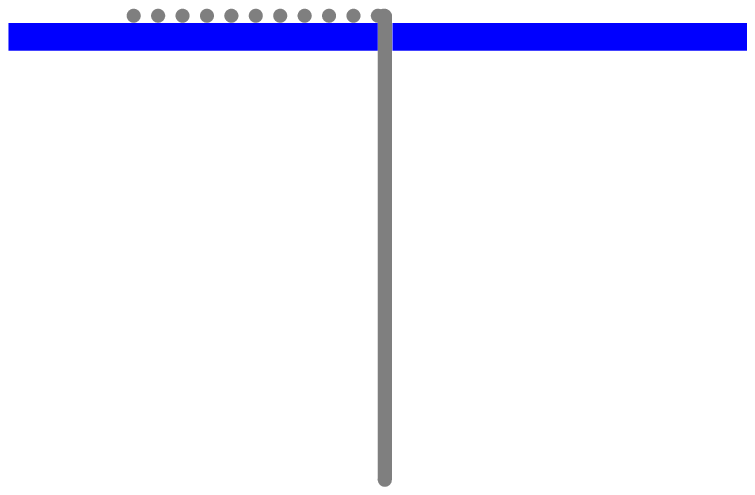}}
  \caption{Sketches for the three chain problems treated here. (a) Falling chain: the straight up line represents the instantaneous configuration of a chain falling vertically on a smooth, rigid plane, where dots represent the accumulating links at rest. (b) Folded chain: a chain is folded about a sharp bight, with one arm attached to a fixed point and the other arm supposed to remain vertical during its downward motion. (c) Sliding chain: a chain is sliding through a hole in a smooth, rigid plane, being continuously lengthened by the addition of quiescent links (represented as dots) that are abruptly connected with the moving chain. In all these sketches, gravity is directed downward along the vertical line.
}
\label{fig:chain_sketches}
\end{figure}

Specifically, these are the problems we shall treat here (and certainly, not for the first time). Problem (a), or the problem of a \emph{falling chain}: a chain initially kept at equilibrium over a smooth, rigid plane, is let go under the action of its own weight; describe the motion of the chain while the links coming in contact with the supporting plane accumulate over it, and compute the total force impressed on it, that is, the \emph{dynamical weight} of the chain, see Fig.~\ref{fig:chain_sketches}(a).

Problem (b), or the problem of a \emph{folded chain}: a chain is initially held in a U-shaped configuration, with one arm ending in a fixed point and the other free to fall under its own weight; describe the motion until when the chain is fully stretched vertically and compute the dynamical weight of the chain, see Fig.~\ref{fig:chain_sketches}(b).

Problem (c), or the problem of a \emph{sliding chain}: a chain is supported by a smooth, rigid plane and leaves it though a hole under the action of gravity; describe the motion of the chain hanging vertically below the plane and compute the dynamical weight of the chain, see Fig.~\ref{fig:chain_sketches}(c).

\subsubsection{Early History}\label{sec:early_history}
It is hard to say which of these problems is the oldest. For example, a variant of Problem (a) is treated in Tait and Steele's treatise~\cite[p.\,301]{tait:treatise}. Here one end of a uniformly heavy chain hangs over a small, smooth pulley and the other end is coiled up on a table; if the free arm of the chain preponderates the antagonistic arm is pulled upward and keeps removing links from the table; if, on the contrary, the free arm is outweighed by the other arm, this problem becomes very similar to Problem (a).

Problem (b) is treated in Love's book~\cite[p.\,261]{love:theoretical}; here the free chain's arm is supposed to fall with acceleration $g$ and the jump in tension at the tight chain's bight is derived from the localized balance of linear momentum.

Problem (c) featured in both Jean's \cite[p.\,236]{jeans:elementary} and Love's \cite[p.\,260]{love:theoretical} treatises, which arrive at the same equation of motion obtained by Cayley~\cite{cayley:class} from his general variational principle for the motion of systems whose mass grows by continuously taking into ``connextion'' (in Cayley's own language) the contents of a quiescent reservoir. The solution method suggested by those textbooks was based on the balance of linear momentum, and no mention was made there of Cayley's principle, though both methods produced one and the same equation of motion when applied to Problem (c).

In solving both Problems (b) and (c), Love's approach, unlike others, requires computing the tension within the chain.\footnote{The tension is indeed an important player in the dynamics of chain, at least as important as the equation of motion, especially at the points where it is  discontinuous. These latter are admirably described by Love~\cite[p.\,259]{love:theoretical}: ``It often happens
that two parts of a chain move in different ways, and that portions of the chain are continually transferred from the part that is moving in one way to the part that is moving in the other way. The tension at the place where the motion changes is then to be determined by the principle that the increase of momentum of a system in any interval is equal to the impulse of the force which acts upon it during that interval.''} Moreover, also in view of our further development, I found very illuminating the following sentence from Love's book \cite[p.\,260]{love:theoretical}: ``It is important to observe that discontinuous motions such as are considered here in general involve dissipation of energy.''

\subsubsection{Later History}\label{sec:later_history}
Before discussing the details of the solutions to Problems (a)-(c) afforded by the theory proposed in this paper, also to place them in a clearer perspective, I summarize below some other, newer approaches to these problems, some perhaps more convincing than others.

One popular solution to Problem (a) assumes that the chain's links fall with acceleration $g$ and computes the force that they exert on the supporting plane by balancing the linear momentum when they come abruptly at rest \cite{feynman:feynman}. The conclusion is that at each instant in time the dynamical weight of the chain is  three times the static weight of the chain collected on the plane up to then. This solution was already common knowledge by the early 1950s \cite{miller:weight} and also received some experimental confirmation, mixed with the voicing of some concerns as to the validity of the assumption that the chain falls with acceleration $g$, or \emph{freely}, as one might also say \cite{satterly:falling}.\footnote{The assumption of energy conservation was equally questioned in \cite{satterly:falling}.}

It was indeed shown experimentally by Hamm and G\'eminard~\cite{hamm:weight} that a chain does \emph{not} fall freely on a supporting plane, but with an acceleration \emph{greater} than $g$. This naturally led then to envision the existence of a dynamical force pulling the chain down, which can only be exerted by the solid surface on which the chain is accumulating, and for which an explicit representation was also proposed, based on a model for the growth of the accumulating coil \cite{hamm:weight}. As remarked in \cite{grewal:chain}, this force is for all intents and purposes a \emph{tension} exerted by the supporting surface, pulling downward the falling chain.\footnote{The free fall corresponding of course to the case when the pulling tension vanishes.} As a result, the dynamical weight of the chain now exceeds three times the static weight of the accumulated chain (even dramatically so, as the experimental figures on \cite{hamm:weight} suggest that it is nearly 6 times the total static weight of the chain at the time immediately preceding its complete deposition). In Section~\ref{sec:falling} below, we shall make a more definite contact with both this study and that of Grewal, Johnson and Ruina~\cite{grewal:chain}, which are equally central to our development.\footnote{I believe that they were also at the basis of the explanation given by Biggins and Warner~\cite{biggins:understanding,biggins:growth} for the \emph{chain fountain} problem.}

Problem (b) is perhaps the most controversial of the three considered here. If, as we learned above, Love~\cite[p.\,261]{love:theoretical} assumed that the chain's arm with a free end falls freely,\footnote{So did also Lamb~\cite[p.\,149]{lamb:dynamics}, as recounted in \cite{calkin:dynamics_I}.} Hamel~\cite[p.\,643]{hamel:theoretische}, as before him had done Kucharski~\cite{kucharski:kinetik}, rejected this assumption and based instead the solution of the problem of the folded chain on the conservation of the total mechanical energy.\footnote{This is also the avenue taken in \cite{wong:falling}, albeit phrased within a purely Lagrangean approach.} This latter implies that the velocity of the falling arm increases indefinitely as the chain approaches the fully stretched configuration. This can be easily understood by realizing that conservation of energy requires a finite amount of potential energy to be transferred to a moving chain fragment whose mass decreases to nought as the chain straightens to its full length. Direct inspection of the equation of motion derived from the energy conservation shows that also the acceleration diverges in the same limit, and so the chain is bound to fall faster than in a free fall.\footnote{The energy conserving solution to Problem (b) was also mentioned by Hagedorn~\cite{hagedorn:some}, who found it instrumental to illustrate the need for introducing stability criteria for continuous systems that measure discrepancies ``in the mean'', instead of pointwise. A pointwise stability measure would indeed reckon unstable the vertical downward equilibrium configuration of a chain in a uniform gravitational field, as had Singh and Denim~\cite{singh:about} found the straight, centripetal configuration of a heavy flexible string attached to a satellite in a central field of attraction.}

Again a positive tension, $\taup$, applied to the falling arm at the chain's bight must be responsible for such a faster fall. It was realized in \cite{calkin:dynamics_I} that conservation of energy actually requires that $\taup$ equals the tension, $\taum$, acting at the bight on the resting arm. As also pointed out in \cite{schagerl:paradox,schagerl:dymamics}, different \emph{constitutive} choices for these tensions would grant a full host of evolutions for the folded chain, including the one where the free arm falls freely \cite{steiner:equations}. The problem, however, is that we are not at liberty of choosing constitutive laws for $\taup$ and $\taum$, at least as long as we model chains as \emph{inextensible} strings, for which the tension in an internal \emph{reactive} force to be determined. We shall show in Section~\ref{sec:folded} how this task can be accomplished within our theory.

The chain problems treated here have been selected with the purpose of proving that, treated in the appropriate continuum theory, they are free from any paradox. Studies that treat the problem of a folded chain with the methods and language appropriate to a discrete systems have also been pursued; among these, two in particular deserve notice \cite{tomaszewski:dynamics,tomaszewski:motion}.

In the classical solution of Problem (c), when the chain starts sliding downward from rest and zero dangling length, the motion is predicted to take place with uniform acceleration $a=\frac13g$. Although nothing would of course be paradoxical with such a prediction, echoing Love~\cite[p.\,260]{love:theoretical}, Sommerfeld~\cite[p.\,257]{sommerfeld:mechanics} remarked that in the corresponding motion the total mechanical energy is not conserved and made Carnot's energy loss mechanism responsible for this \cite[pp.\,28, 29, 241]{sommerfeld:mechanics}. Such a conclusion was rejected in \cite{wong:falling} and \cite{wong:falling_Hopkins} in favour of a Lagrangean theory that enforces energy conservation. Thus, if not paradoxical, Problem (c) has become at least controversial, even more so on account of the fact that a convincing  explanation for the energy loss can be given in terms of the (inelastic) impacts suffered by the links continuously set in motion by the sliding chain.\footnote{A detailed computation to this effect can be found in \cite{keiffer:falling} and, as remarked in \cite{chicon:comment}, a similar mechanism is also described in \cite[pp.\,25, 28]{saletan:theretical} and \cite[p.\,46]{jose:classical}.}

In the following three sections we shall study the chain problems described in this section within the general theory proposed here. The only constitutive parameter will be $f$ in \eqref{eq:W_s_formula_chain}. Whenever $f>0$, the singular point which will be appropriately identified for the three problems, is the site of a dissipative internal shock. Further sources of dissipation may arise from external shocks, if present; for them \eqref{eq:external_shock_balance_energy_plus_minus} expresses in this theory the analogue of what in Sommerfeld's language \cite{sommerfeld:mechanics} is Carnot's energy loss.

\section{Falling Chain}\label{sec:falling}
In this section and in the following two we solve Problems (a), (b), and (c), respectively, by applying the theory put forward in Sects.~\ref{sec:dissipation} and \ref{sec:shocks}. In particular, we shall make use of equation \eqref{eq:balance_equation_linear_momentum} to describe the motion of a regular arc in a chain, of \eqref{eq:evolution_equation_v_0} to balance the linear momentum at a singular point, and of \eqref{eq:dissipation_gold}, which combined with \eqref{eq:evolution_equation_v_0} embodies the balance of energy at a singular point. By \eqref{eq:W_s_formula_chain}, \eqref{eq:dissipation_gold} here acquires the form
\begin{equation}\label{eq:jump_tension}
\jtau=-\frac12 f\lambda\sgn(\sspeed)\jvel^2.
\end{equation}
All the above equations, which are dynamical in nature, must be combined with the condition for kinematic compatibility in \eqref{eq:jump_condition_velocity}.

\begin{figure}[!h]
  \centering
  \hspace{0.05\linewidth}
  \includegraphics[width=.3\linewidth]{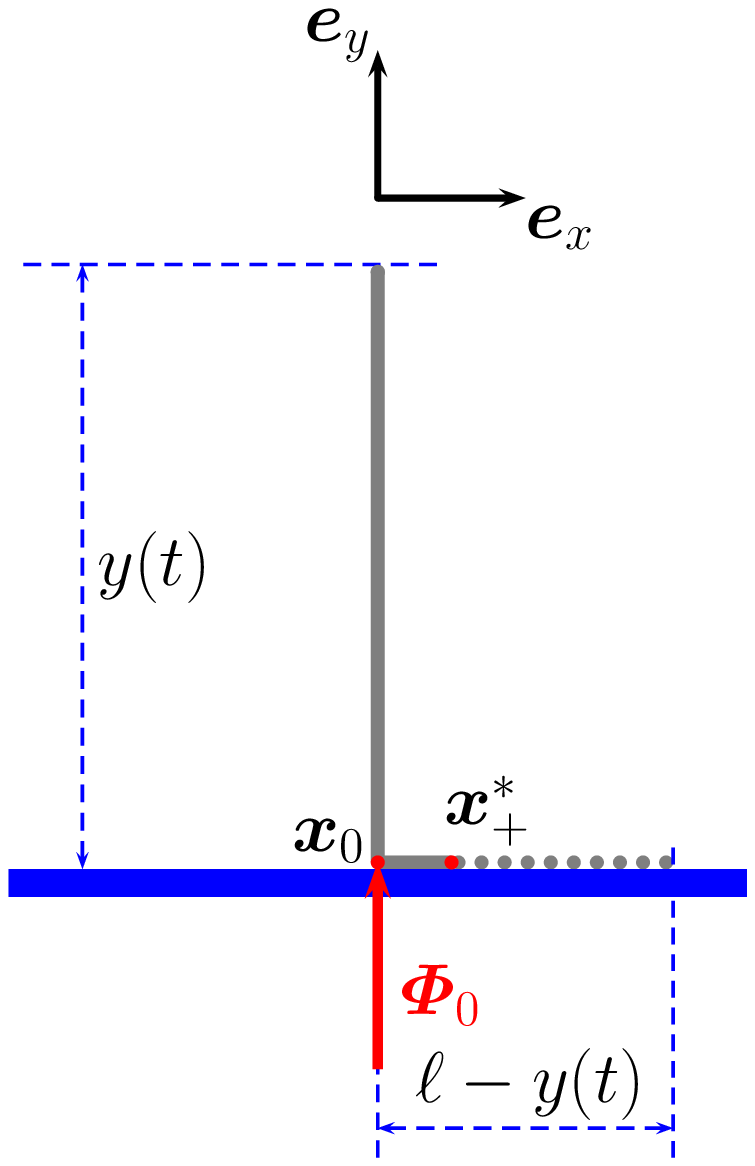}
  \hspace{0.10\linewidth}
  \includegraphics[width=.5\linewidth]{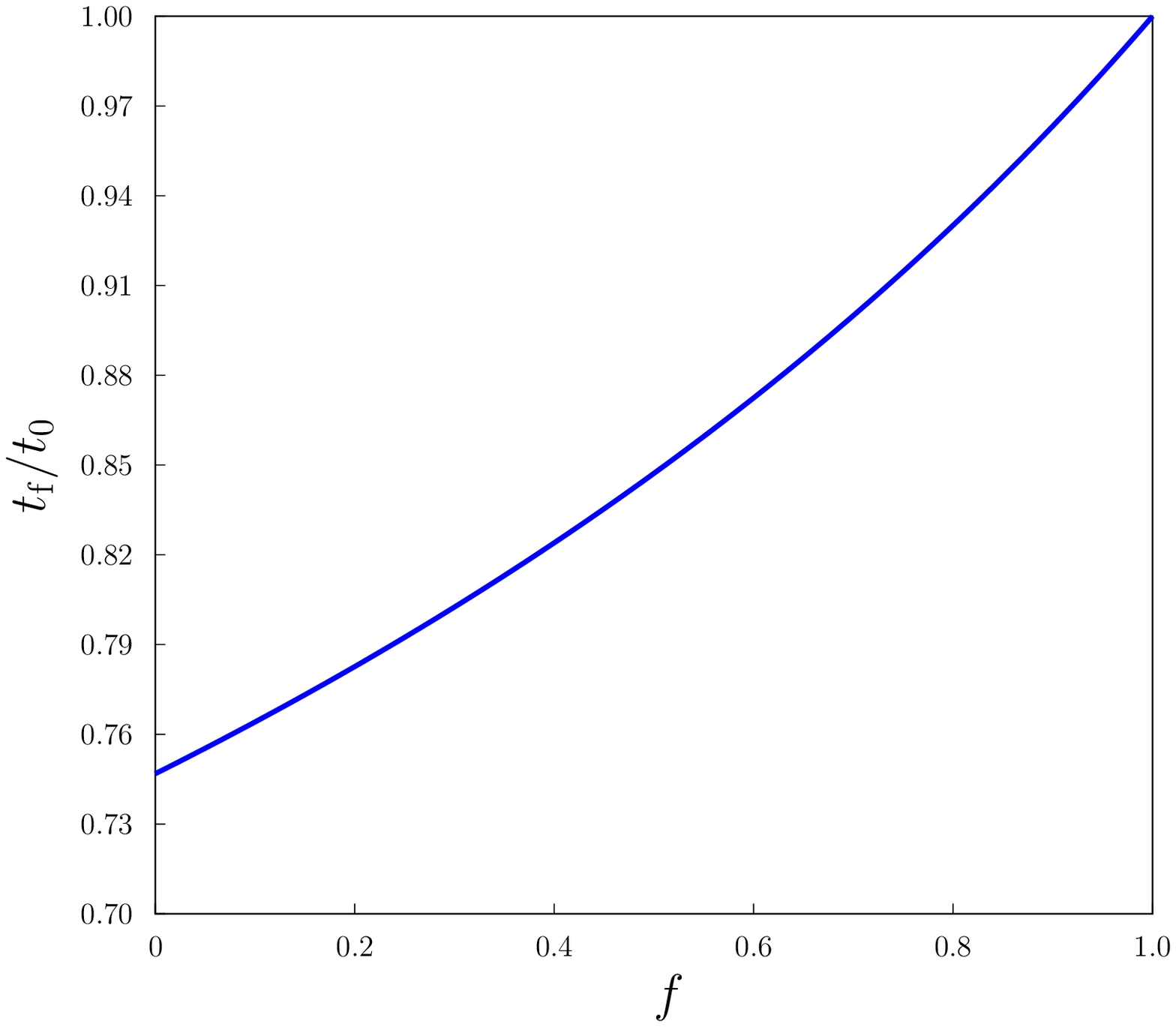}
  \caption{(a) Decorated sketch of our model for the dynamics of a falling chain. Two singular points are located at $\xnot$ and $\xstarp$, the former being an internal shock and the latter an external one. The height of the falling chain at time $t$ is $y(t)$; the length being deposited on the supporting plane at the same time is $\ell-y(t)$, where $\ell$ is the total length of the chain. $\bm{\Phi}_0$ is the reactive, impulsive force continuously imparted by the plane to the chain at $\xnot$. (b) The time of fall $\tf$ scaled to $t_0$ in \eqref{eq:falling_chain_t_0} as a function of the parameter $f$.
}
\label{fig:falling_chain}
\end{figure}
Figure~\ref{fig:falling_chain}(a), which is a decorated version of the sketch in Fig.~\ref{fig:chain_sketches}(a), describes our idealized model for a falling chain. Here $\xnot$ denotes the point where the chain comes in contact with a smooth, rigid plane; we think of it as a singular point for the chain dynamics, where the unit tangent $\ut$ jumps from $\tm=-\ey$ to $\tp=\ex$ in the Cartesian frame $(\ex,\ey)$ at the value $s=s_0(t)$ of the arc-length parameter that describes the material points in the chain with origin in its free end. An internal shock in propagating at $\xnot$ with speed $\sspeed$. At the point $\xstarp$, which we shall imagine to be arbitrarily close to $\xnot$, the chain comes in contact with the coil of links already deposited on the supporting plane. This is the site of an external shock, governed by (the plus side of) equations \eqref{eq:external_shock_plus_minus}.

Here we study the dynamics of this chain in the special class of motions that preserve its shape, characterized by \eqref{eq:motion_along_its_length}. By the inextensibility constraint, at instant $t$ in time $v(t)$ is the velocity of all links in the moving chain. The compatibility condition \eqref{eq:jump_condition_velocity} indeed readily gives that
\begin{equation}\label{eq:falling_chain_kinematic_compatibility}
v^+=v^-=v=-\sspeed,
\end{equation}
and so the internal shock is standing in space, as also is the external shock at $\xstarp$, by \eqref{eq:external_shock_speeds}.

The theory we adopt does not require any specific assumption on the shape of the quiescent coil deposited
on the plane. We only need to know the force $-\Phistarp$ it suffers from the chain. Since we assume $\xstarp$ to be arbitrarily close to $\xnot$, we identify the tension $\taustarp$ featuring in \eqref{eq:external_shock_balance_linear_momentum_plus_minus} with the right limit $\taup$ of the tension in the chain at $\xnot$ and we write \eqref{eq:external_shock_balance_linear_momentum_plus_minus} as
\begin{equation}\label{eq:falling_chain_external_shock_balance_linear_momentum}
\Phistarp=(\taup-\lambda v^2)\ex.
\end{equation}

At $\xnot$ the chain is also subject to the linear momentum supply $\bm{\Phi}_0$, which here plays the role of $\Force$ in \eqref{eq:evolution_equation_v_0} and is taken to be orthogonal to the supporting plane (since the latter is smooth), $\bm{\Phi}_0=\Phi_0\ey$, with $\Phi_0\geqq0$. Having a reactive nature, $\Phi_0$ is to be determined along with the motion. Writing \eqref{eq:evolution_equation_v_0} in the frame $(\ex,\ey)$, we readily conclude that
\begin{subequations}\label{eq:falling_chain_singular_momentum_balance}
\begin{align}
\taup&=\lambda v^2,\label{eq:falling_chain_tension_plus}\\
\Phi_0&=\lambda v^2-\taum,\label{eq:falling_chain_Phi_not}
\end{align}
\end{subequations}
where $\taum$ is the left limit of the tension at $\xnot$, responsible when positive for a faster fall of the chain. Making use of \eqref{eq:falling_chain_tension_plus} in \eqref{eq:falling_chain_external_shock_balance_linear_momentum}, we see that $\Phistarp=\zero$, and so the coil of accumulating links is free from any horizontal forces. To determine $\Phi_0$, we need to know $\taum$, which since $\Phi_0\geqq0$ is subject to the bound
\begin{equation}\label{eq:falling_chain_tau_minus_bound}
\taum\leqq\lambda v^2.
\end{equation}

If, as in Fig.~\ref{fig:falling_chain}(a), $y(t)$ denotes the hight of the falling chain at time $t$ and $\ell$ is the total length of the chain, then $\ell-y(t)$ is the length of the chain accumulated on the plane. Correspondingly, $v=-\dot{y}$, where a superimposed dot denotes differentiation with respect to time, and \eqref{eq:falling_chain_kinematic_compatibility} becomes
\begin{equation}\label{eq:falling_chain_shock_speed}
\dot{y}=\sspeed.
\end{equation}
Since along the falling chain $\ba=\ddot{y}\ey$ and $\force=\lambda g\ey$, equation \eqref{eq:balance_equation_linear_momentum} yields
\begin{equation}\label{eq:falling_chain_tension_pde}
\ps\tau=-\lambda(\ddot{y}+g)\quad\text{for}\quad 0\leqq s\leqq y.
\end{equation}
Being the left side of equation \eqref{eq:falling_chain_tension_pde} independent of $s$ and  $\tau=0$ at the free end $s=0$, it readily follows that
\begin{equation}\label{eq:falling_chain_tau_minus}
\taum=-\lambda (\ddot{y}+g)y.
\end{equation}
Since $\jvel^2=2\dot{y}^2$ at $\xnot$, combining \eqref{eq:falling_chain_tau_minus}, \eqref{eq:falling_chain_shock_speed}, \eqref{eq:falling_chain_tension_plus}, and \eqref{eq:jump_tension}, we obtain the equation
\begin{equation}\label{eq:falling_chain_equation_motion_sgn}
\left(1+f\sgn(\ydot)\right)\ydot^2+(\yddot+g)y=0,
\end{equation}
which actually describes the motion of the chain only in terms of the constitutive parameter $f$.

Shortly below we shall give an analytic representation of certain solutions of \eqref{eq:falling_chain_equation_motion_sgn}. Some of their qualitative features can be easily obtained directly from \eqref{eq:falling_chain_equation_motion_sgn}. First, if $0\leqq f<1$, then $\yddot+g<0$, and so the chain falls with an acceleration greater than $g$. Second, if $0\leqq f\leqq1$ and $\ydot(0)=0$, which is when the chain is released with initial zero velocity, then $\sgn(\ydot)\equiv-1$ all along the motion, and so \eqref{eq:falling_chain_equation_motion_sgn} becomes
\begin{equation}\label{eq:falling_chain_equation_motion}
(1-f)\ydot^2+(\yddot+g)y=0,
\end{equation}
which coincides with (5) of \cite{hamm:weight}, provided that $f$ is set equal to the parameter $\gamma$ introduced there.

Accordingly, by \eqref{eq:falling_chain_shock_speed}, \eqref{eq:falling_chain_tension_plus}, and \eqref{eq:jump_tension}, we can write the tension $\taum$ exerted on the falling chain at $\xnot$ as
\begin{equation}\label{eq:falling_chain_tension_minus}
\taum=(1-f)\lambda\ydot^2,
\end{equation}
which obeys \eqref{eq:falling_chain_tau_minus_bound} and vanishes for $f=1$, that is, when the chain is in free fall. By inserting \eqref{eq:falling_chain_tension_minus} into \eqref{eq:falling_chain_Phi_not}, we arrive at
\begin{equation}\label{eq:falling_chain_Phi_not_explicit}
\Phi_0=f\lambda\ydot^2,
\end{equation}
which agrees with (6) of \cite{hamm:weight}. Moreover, the tension $\taum$ was denoted $N_1$ in \cite{grewal:chain} and we may identify with $\Phi_0$ the force denoted $N_2$ there. By comparing \eqref{eq:falling_chain_tension_minus} and \eqref{eq:falling_chain_Phi_not_explicit} we can easily express our $f$ in terms of $N_1$ and $N_2$,\footnote{I assume that there is a typo in the formula expressing $\gamma$ in terms of $N_1$ and $N_2$ in \cite[p.\,728]{grewal:chain}. That formula would be reconciled with \eqref{eq:falling_chain_f_Ns} by exchanging $N_1$ and $N_2$.}
\begin{equation}\label{eq:falling_chain_f_Ns}
f=\frac{N_1}{N_1+N_2},
\end{equation}
an identification which is further confirmed by comparing \eqref{eq:falling_chain_equation_motion} with (10) of \cite{grewal:chain}. The advantage of \eqref{eq:falling_chain_tension_minus} and \eqref{eq:falling_chain_Phi_not_explicit} over \eqref{eq:falling_chain_f_Ns} is that it expresses both $N_1$ and $N_2$ in terms of a single constitutive parameter, $f$. Thus, for the record, the special chains ingeniously constructed in \cite{grewal:chain} would correspond to $f=\frac56$ and $\frac12$, see equations (18) and (19) of \cite{grewal:chain}, respectively.

$\Phi_0$ is also the dynamic weight of the chain that is yet to fall; to obtain the total dynamic weight $P$ of the falling chain we need to add to $\Phi_0$ the static weight of the chain that has already fallen:
\begin{equation}\label{eq:falling_chain_P_first}
P=f\lambda\ydot^2+\lambda g(\ell-y).
\end{equation}
Once \eqref{eq:falling_chain_equation_motion} is solved for $y(t)$, \eqref{eq:falling_chain_P_first} shall deliver $P$ as a function of time.

Hamm and G\'eminard~\cite{hamm:weight} used $f$ (which they called $\gamma$), as a fitting parameter to explain their measurements indicating that a chain was indeed falling with acceleration greater than $g$. They determined $f\approx0.83$ and, more importantly, they found that this value was independent of the material properties of the floor on which the chain was falling, supporting our view that $f$ is indeed a constitutive parameter of the chain only.

The (implicit) solution of \eqref{eq:falling_chain_equation_motion} can be found by quadratures. To free the solution from inessential physical constants, we rescale both lengths and time. Lengths will be rescaled to the total length $\ell$ of the chain. Time will be rescaled to
\begin{equation}\label{eq:falling_chain_t_0}
t_0:=\sqrt{\frac{2\ell}{g}},
\end{equation}
which is the time of \emph{free fall} from the height $\ell$. Thus, letting
\begin{equation}\label{eq:falling_chain_eta_xi}
\eta:=\frac{y}{\ell}\quad\text{and}\quad\xi:=\frac{t}{t_0},
\end{equation}
equation \eqref{eq:falling_chain_equation_motion} is transformed into
\begin{equation}\label{eq:falling_chain_equation_motion_eta}
(1-f){\eta'}^2 +(\eta'' +2)\eta=0,
\end{equation}
where a prime $'$ denotes differentiation with respect to $\xi$. Correspondingly, letting $P_0:=\lambda\ell g$ be the total static weight of the chain, we obtain from \eqref{eq:falling_chain_P_first} that
\begin{equation}\label{eq:falling_chain_P_rescaled}
\frac{P}{P_0}=\frac12f{\eta'}^2+1-\eta.
\end{equation}

\begin{figure}[!h]
  \centering
  \subfigure[]{\includegraphics[width=.48\linewidth]{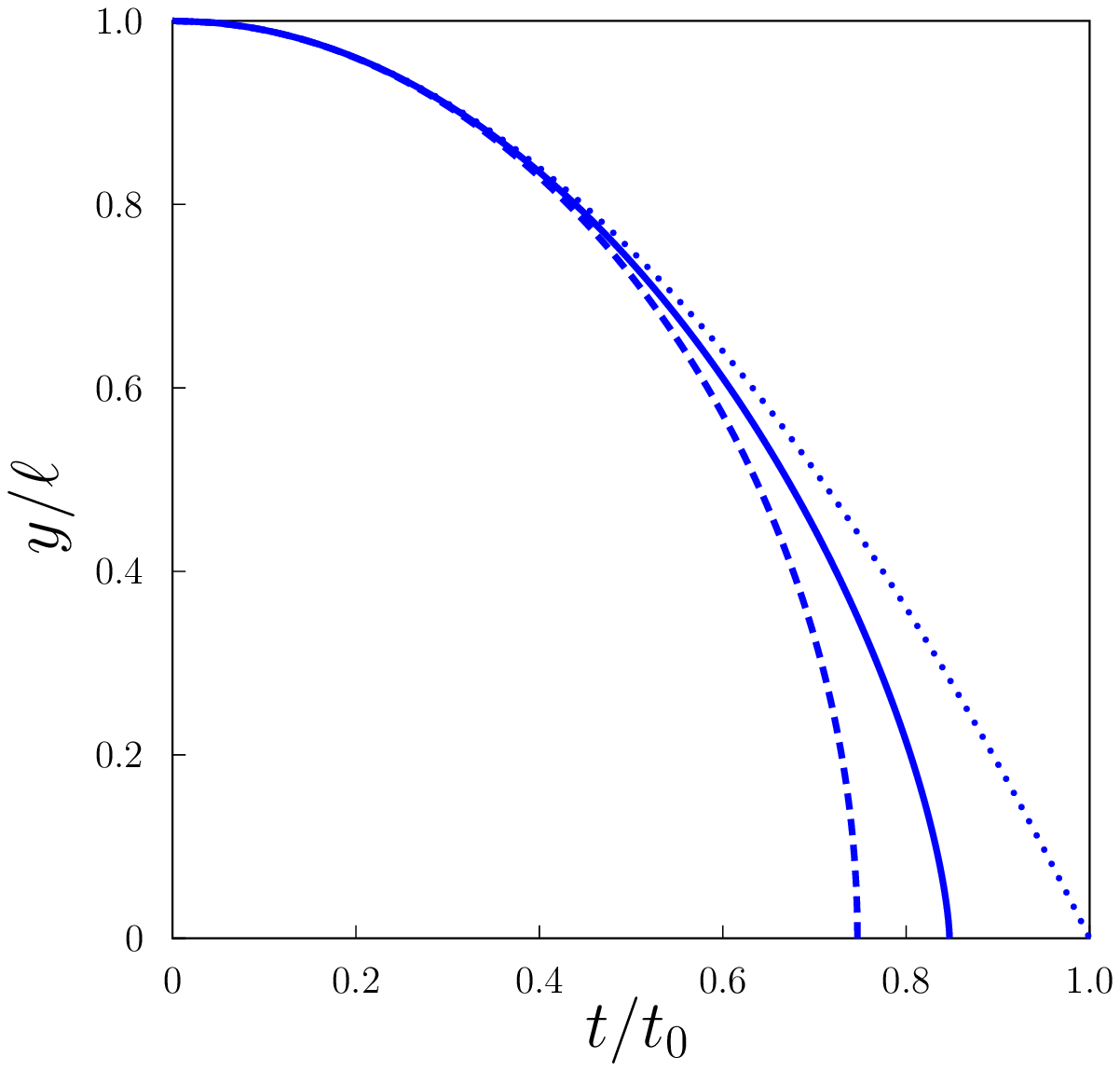}}
  \hspace{0.02\linewidth}
  \subfigure[]{\includegraphics[width=.48\linewidth]{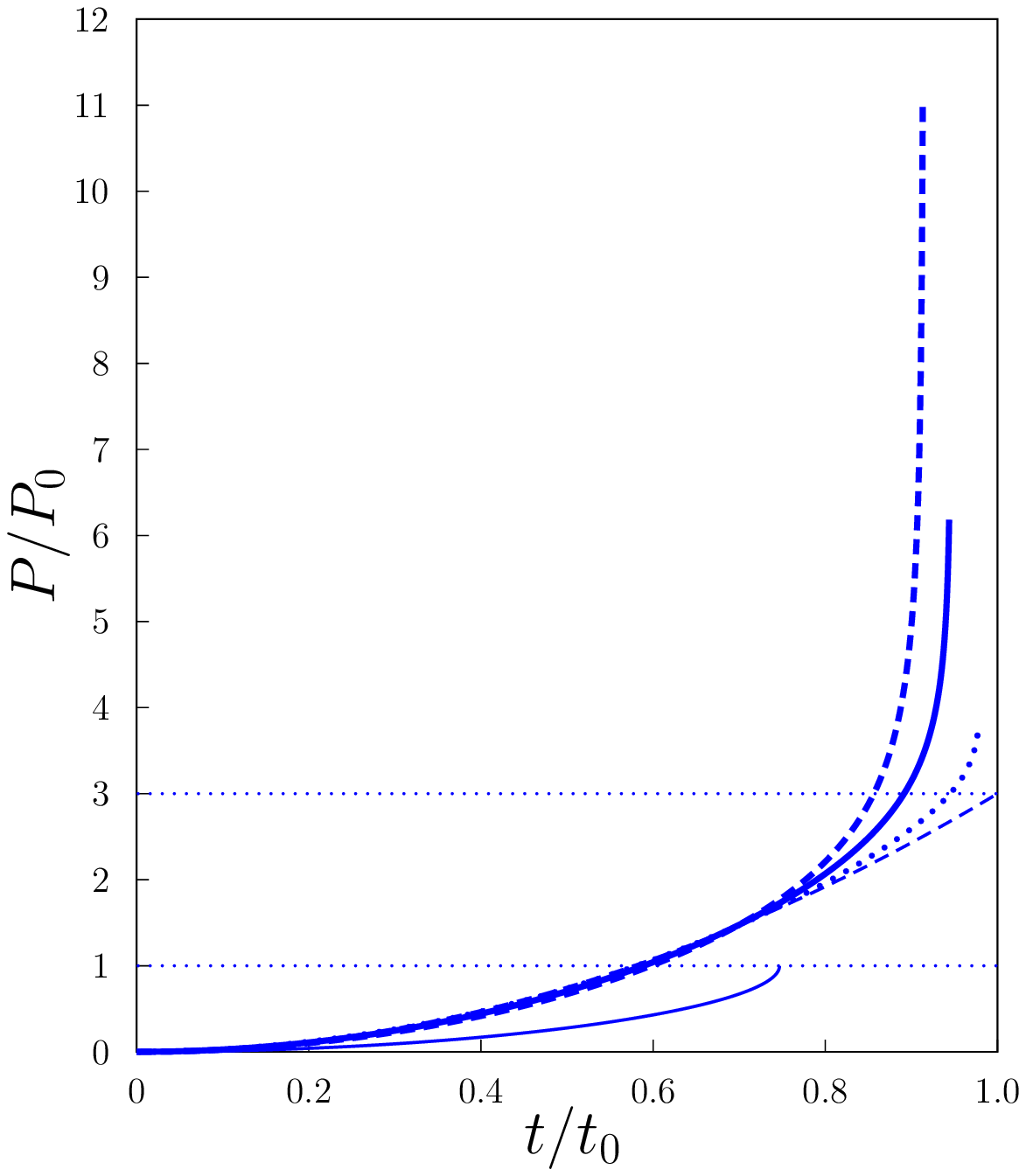}}
  \caption{(a) Plots of $y$ scaled to $\ell$ as a function of the scaled time $t/t_0$ for $f=1$ (dotted line), $f=\frac12$ (solid line), and $f=0$ (dashed line). (b) The dynamic weight $P$ of the falling chain scaled to the total static weight $P_0$ as a function of $t/t_0$ for $f=0$ (this solid line), $f=0.75$ (dashed line), $f=0.85$ (solid line), $f=0.95$ (dotted line), and $f=1$ (thin dashed line). Each graph is plotted over the time interval during which the solution to \eqref{eq:falling_chain_equation_motion_eta} in \eqref{eq:falling_chain_xi_of_eta} reaches $\eta=\varepsilon$ decreasing from $\eta=1$. For all graphs, $\varepsilon=0.01$.
}
\label{fig:falling_chain_y_P_plots}
\end{figure}

The solution of \eqref{eq:falling_chain_equation_motion_eta} under the initial conditions $\eta(0)=1$ and $\eta'(0)=0$ reads as
\begin{equation}\label{eq:falling_chain_xi_of_eta}
\xi=\frac12\sqrt{3-2f}\int_\eta^1\frac{1}{\sqrt{x^{2(f-1)}-x}}dx,
\end{equation}
whence it follows that the \emph{time of fall} $\tf$ is given by
\begin{equation}\label{eq:falling_chain_time_of_fall}
\frac{\tf}{t_0}=\frac12\sqrt{3-2f}\int_0^1\frac{1}{\sqrt{x^{2(f-1)}-x}}dx,
\end{equation}
which is function of $f$ only.
Figure~\ref{fig:falling_chain}(b) shows the graph of this function; it is apparent that $\tf\leqq t_0$, with equality holding only for $f=1$. It also makes us appreciate quantitatively how faster than freely a chain may actually fall. Figure~\ref{fig:falling_chain_y_P_plots}(a) illustrates the graphs of $\eta$ against $\xi$ according to \eqref{eq:falling_chain_xi_of_eta} for three values of $f$.
They indicate that, apart from the case where $f=1$, $\ydot$ diverges to $-\infty$ as $t\to\tf$, which indeed follows directly from \eqref{eq:falling_chain_xi_of_eta}, as also does the similar divergence of $\yddot$ in the same limit. As a consequence, by \eqref{eq:falling_chain_P_rescaled}, $P$ is predicted to diverge to $+\infty$ as $t\to\tf$. By \eqref{eq:falling_chain_xi_of_eta}, equation \eqref{eq:falling_chain_P_rescaled} can be given an easier expression,
\begin{equation}\label{eq:falling_chain_P_eta_only}
\frac{P}{P_0}=1+\frac{2f\eta^{2(f-1)}-3\eta}{3-2f},
\end{equation}
which shows how $P$ diverges as $\eta\to0$, unless $f=1$ or $f=0$. The former is the case of free fall, when \eqref{eq:falling_chain_P_eta_only} reduces to $P=3P_0(1-\eta)$. The latter is the case of no dissipation at the internal shock, when \eqref{eq:falling_chain_P_eta_only} reduces to $P=P_0(1-\eta)$.\footnote{If the behaviour of $P$ for $f=0$ appears a bit peculiar, we shall offer below an independent argument that would suggest that $f$ should indeed be greater than or equal to $\frac12$.}

The unboundedness of $\ydot$, $\yddot$, and $P$ could be regarded as paradoxical. It is indeed a consequence of the continuum model being used here, which can hardly be considered valid up to when $y$ vanishes.\footnote{A further limitation of the model is the strict one-dimensionality assumed for the motion, which is likely to be violated as $y\to0$.} Following \cite{wong:falling} and \cite{geminard:motion}, we introduce a cutoff length, $\ell_0:=\varepsilon\ell$, which could be estimated as consisting of up to 3 chain's links \cite{geminard:motion}, so that $\eta$ would be subject to the bound $\eta\geqq\varepsilon$. Such a bound alters only marginally the time of fall $\tf$, especially if $\varepsilon$ is sufficiently small, but it makes the limiting value of $P$ finite. Figure~\ref{fig:falling_chain_y_P_plots}(b) illustrates the graphs of $P$ against time for several values of $f$; they are drawn up to the time when $\eta=\varepsilon=0.01$.

We close this section by computing the total dissipation associated with our solution of the falling chain problem. There are two independent sources of dissipation in this system, one for each shock. At the internal shock, by \eqref{eq:W_s_formula_chain} and \eqref{eq:falling_chain_shock_speed}, the dissipative power $\Ws$ is given by
\begin{subequations}
\begin{equation}\label{eq:falling_chain_Ws}
\Ws=-f\lambda v^3=f\lambda\ydot^3.
\end{equation}
At the external shock, by \eqref{eq:external_shock_balance_energy_plus_minus} and \eqref{eq:falling_chain_tension_plus}, the dissipative power $\Wstarp$ is given by
\begin{equation}\label{eq:falling_chain_W_star_plus}
\Wstarp=\frac12\lambda v^3=-\frac12\lambda\ydot^3.
\end{equation}
\end{subequations}
That both these expressions are compatible with the total balance of energy is seen by computing the time rates of the total kinetic energy
\begin{equation}\label{eq:falling_chain_K}
K=\frac12\lambda y\ydot^2
\end{equation}
and the total potential energy
\begin{equation}\label{eq:falling_chain_V}
V=\frac12\lambda gy^2.
\end{equation}
Making use of \eqref{eq:falling_chain_equation_motion} it is a simple matter to show that
\begin{equation}\label{eq:falling_chain_total_energy_rate}
\dot{K}+\dot{V}=\lambda\left(f-\frac12\right)\ydot^3=\Ws+\Wstarp,
\end{equation}
which coincides with (13) of \cite{hamm:weight}. Since $\ydot\leqq0$, \eqref{eq:falling_chain_total_energy_rate} requires that for the total dissipation to be non-negative (which is the same as requiring that $\Ws+\Wstarp\leqq0$), $f$ must obey the inequality $f\geqq\frac12$.

\section{Folded Chain}\label{sec:folded}
Precisely the same strategy adopted in Section~\ref{sec:falling} to solve the problem of a falling chain is followed here to solve the problem of a folded chain. To avoid unnecessary repetitions, I shall only outline the major steps of the solution and comment the results.

\begin{figure}[!h]
  \centering
  \hspace{0.05\linewidth}
  \subfigure[]{\includegraphics[width=.3\linewidth]{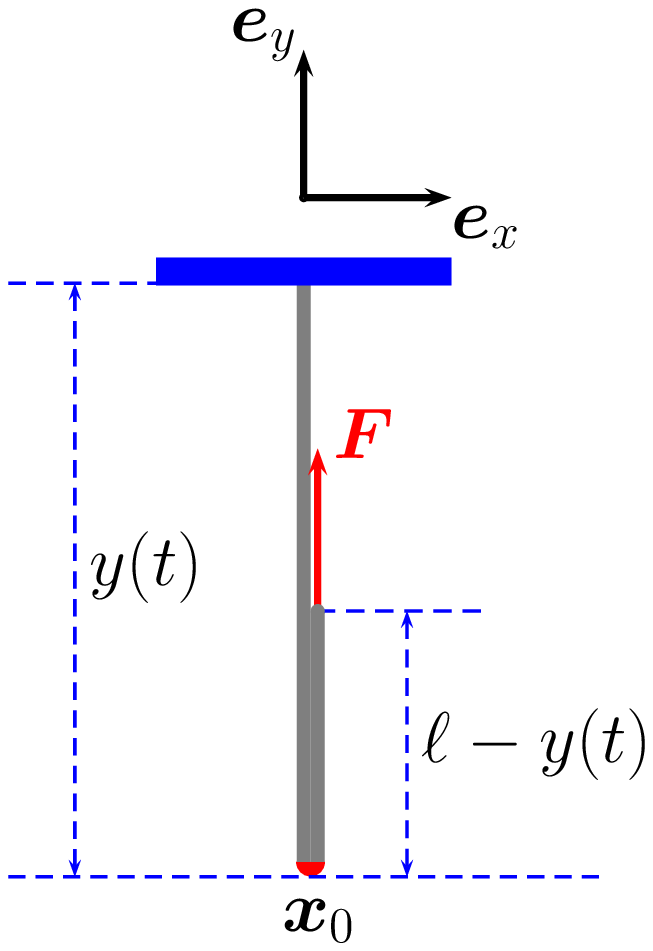}}
  \hspace{0.10\linewidth}
  \subfigure[]{\includegraphics[width=.5\linewidth]{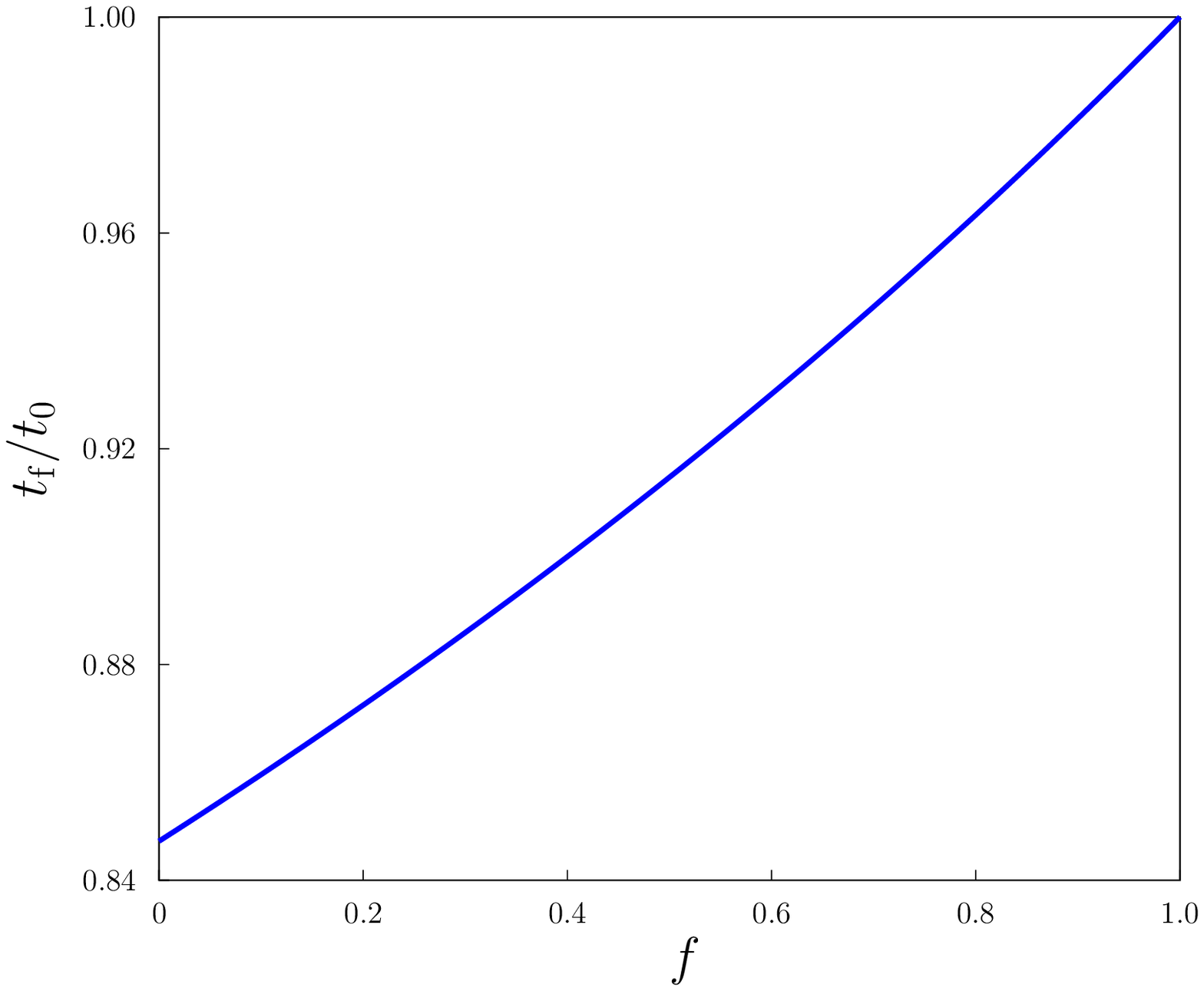}}
  \subfigure[]{\includegraphics[width=.45\linewidth]{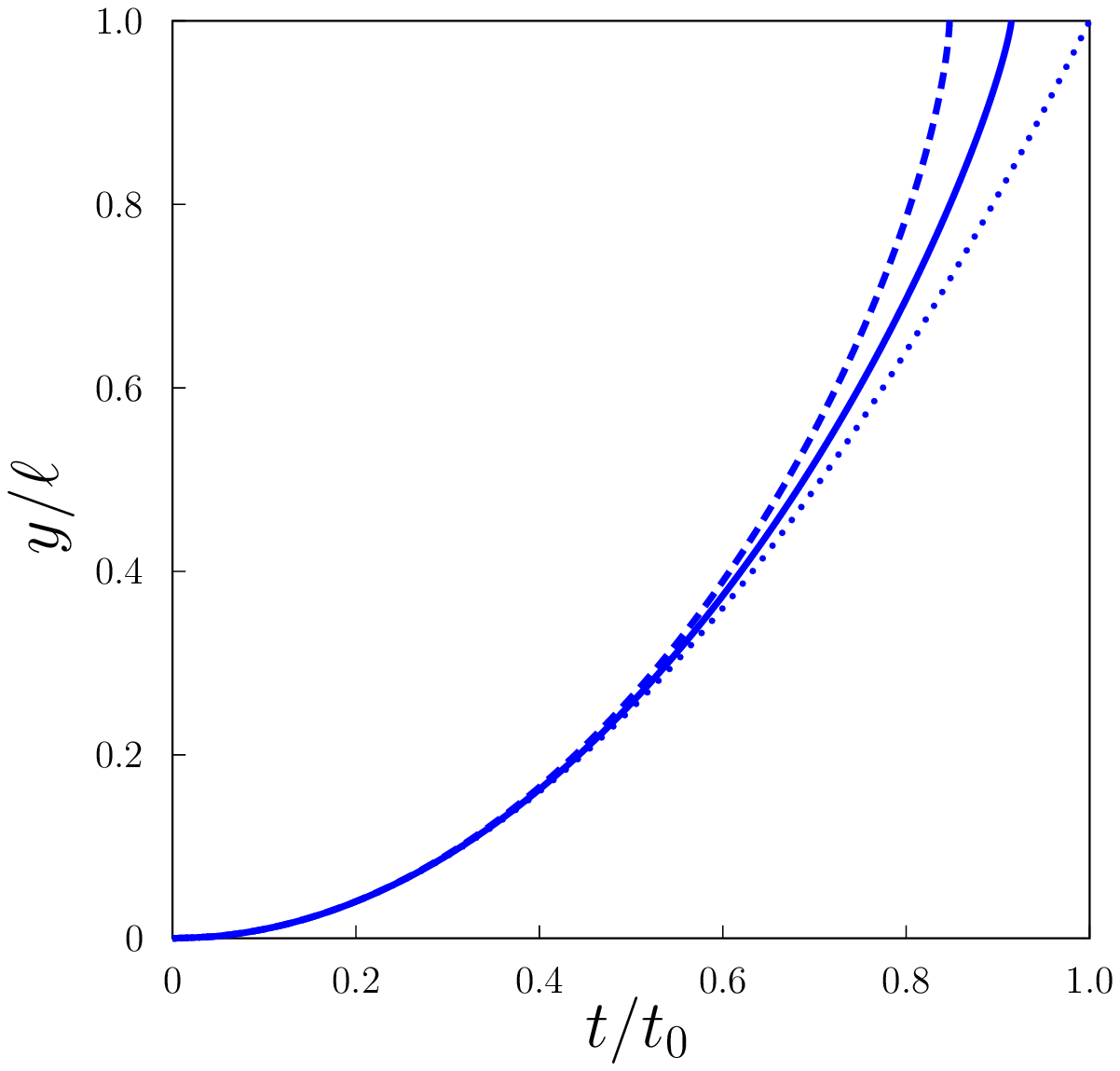}}
  \hspace{0.05\linewidth}
  \subfigure[]{\includegraphics[width=.45\linewidth]{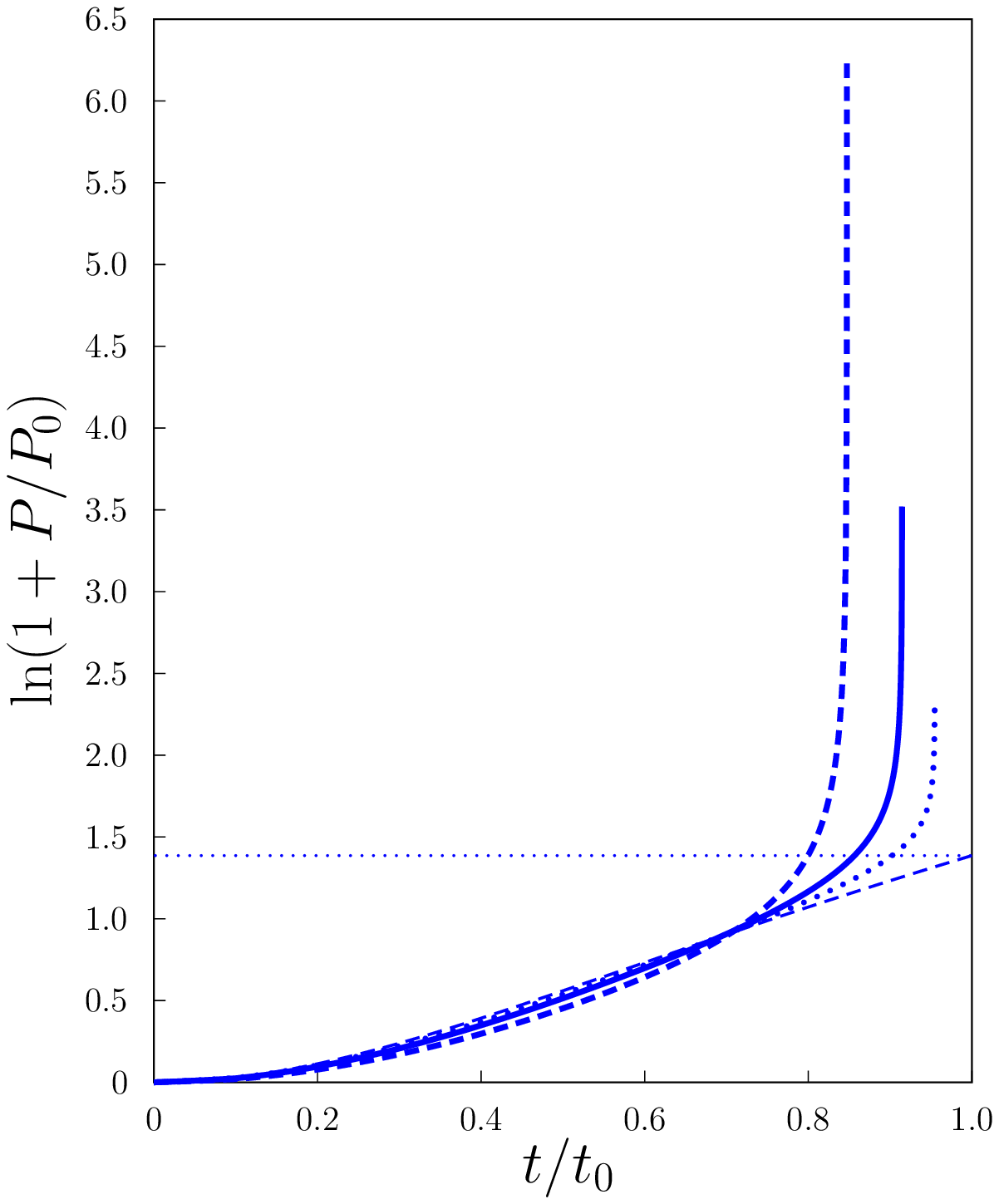}}
  \caption{(a) Ideal model of a folded chain. The singular point $\xnot$, which hosts an internal shock, is the tight bight of the chain. A force $\bm{F}=F\ey$, with $F\geqq0$, is applied at the free end. The origin of the arc-length parameter $s$  is chosen at the fixed point, so that at the free point $s=\ell$, with $\ell$ the total length of the chain. At time $t$, $y(t)$ denotes the length of the arm at rest. (b) The time of fall $\tf$ (scaled to the free fall time $t_0$) is plotted against $f$, according to \eqref{eq:folded_chain_tf}. (c) The graphs of $y$ (scaled to $\ell$) are plotted against the time $t$ (scaled to $t_0$) under the conditions $y(0)=0$ and $\ydot(0)=0$, for $f=0$ (dashed line), $f=0.5$ (solid line), and $f=1$ (dotted line). (d) The logarithm graphs of the dynamic weight $P$ (scaled to the total static weight $P_0$) against time (scaled to $t_0$), for $f=0$ (dashed line), $f=0.5$ (solid line), $f=0.75$ (dotted line), and $f=1$ (thin dashed line). Each graph is plotted in the time interval where $y$ grows up to $(1-\varepsilon)\ell$, with $\varepsilon=0.01$. For $f=1$, the moving arm falls freely and $P$ attains the value $\Pf=3P_0$ at the end of the fall, as confirmed by the corresponding graph, since $\ln4\doteq1.386$.
}
\label{fig:folded_chain}
\end{figure}

Figure~\ref{fig:folded_chain}(a) illustrates the problem in the notation introduced in our theory.
The length of the fallen chain is denoted by $y$; $\ell$ is the total length of the chain, so that the distance between fixed and free ends is $2y-\ell$, and $\xnot$ is the singular point where the chain is tightly folded. As in \cite{steiner:equations} and \cite{reilly:treatment}, we further assume that a vertical force $\bm{F}=F\ey$, with $F\geqq0$, is applied to the free end of the chain. The motion is assumed to take place under gravity and along the vertical line, so that only $y(t)$ suffices to describe it.

Taking the origin of the arc-length parameter $s$ in the fixed point of the chain, we denote by $s_0(t)$ the value of $s$ designating the front of the shock. There, as $s$ increases, $\ut$ jumps from $\tm=-\ey$ to $\tp=\ey$ in the Cartesian frame $(\ex,\ey)$ shown in Fig.~\ref{fig:folded_chain}(a). The kinematic compatibility condition \eqref{eq:jump_condition_velocity} here gives
\begin{equation}\label{eq:folded_chain_kinematic_compatibility}
v^-=0,\qquad v^+=-2\sspeed,
\end{equation}
which replaces \eqref{eq:falling_chain_kinematic_compatibility}. Since the free end moves with speed $\vp=-2\ydot\ey$, inexensibility and \eqref{eq:folded_chain_kinematic_compatibility} require that \eqref{eq:falling_chain_shock_speed} holds also here, and \eqref{eq:jump_tension} implies that
\begin{equation}\label{eq:folded_chain_jump_tension}
\jtau=-2\lambda f\sgn(\ydot)\ydot^2.
\end{equation}
Setting $F$ equal to the tension at the free end and integrating \eqref{eq:balance_equation_linear_momentum}, we arrive at
\begin{equation}\label{eq:folded_chain_tension_plus}
\taup=F-\lambda(g-\yddot)(\ell-y).
\end{equation}
By \eqref{eq:folded_chain_kinematic_compatibility}, the balance of linear momentum \eqref{eq:evolution_equation_v_0} at $\xnot$ reduces to the equation
\begin{equation}\label{eq:folded_chain_tension_plus+minus}
\taup+\taum=2\lambda\ydot^2,
\end{equation}
which combined with \eqref{eq:folded_chain_tension_plus} and \eqref{eq:folded_chain_jump_tension} delivers
\begin{equation}\label{eq:folded_chain_tension_minus}
\taum=\lambda(1+f\sgn(\ydot))\ydot^2
\end{equation}
and
\begin{equation}\label{eq:folded_chain_equation_motion_y}
\lambda(1-f\sgn(\ydot))\ydot^2+\lambda(g-2\yddot)(\ell-y)=F,
\end{equation}
which is the equation of motion for $y(t)$. For $f=1$, equation \eqref{eq:folded_chain_equation_motion_y} reduces to equations (15) and (16) of \cite{steiner:equations}. Moreover, setting $m=0$ in equation (7.8) of \cite{reilly:treatment}, we recover our equation \eqref{eq:folded_chain_equation_motion_y}, provided that $f$ is identified with the parameter $e$ of \cite{reilly:treatment}.

Since the arm of the chain ending in the fixed point is at rest, the dynamical weight of the whole chain is simply $P=\taum+\lambda gy$, which by \eqref{eq:folded_chain_tension_minus} becomes
\begin{equation}\label{eq:folded_chain_P}
P=\lambda(1+f\sgn(\ydot))\ydot^2+\lambda gy.
\end{equation}

For the folded chain the only source of dissipation is at the internal shock. It is an immediate consequence of \eqref{eq:W_s_formula_chain} and \eqref{eq:folded_chain_jump_tension} that
\begin{equation}\label{eq:folded_chain_Ws}
\Ws=-2\lambda f\sgn(\ydot)\ydot^3,
\end{equation}
which coincides with (7.7) of \cite{reilly:treatment} and agrees with (15) of \cite{crellin:balance} for $f=1$. Moreover, the power $W_0$ expended by the force $\bm{F}$ is easily computed to be
\begin{equation}\label{eq:folded_chain_W0}
W_0=-2F\ydot.
\end{equation}
The balance of total energy requires that
\begin{equation}\label{eq:folded_chain_balance_total_energy}
\dot{K}+\dot{V}=W_0+\Ws.
\end{equation}
Starting from the following expressions for $K$ and $V$,
\begin{equation}\label{eq:folded_chain_K}
K=2\lambda(\ell-y)\ydot^2,
\end{equation}
\begin{equation}\label{eq:folded_chain_V}
V=-\frac12\lambda gy^2-\frac12\lambda g(\ell-y)(3y-\ell),
\end{equation}
it is an easy exercise to show that \eqref{eq:folded_chain_balance_total_energy} is indeed valid along all solutions of \eqref{eq:folded_chain_equation_motion_y}.

A special case of \eqref{eq:folded_chain_equation_motion_y} is worth considering in more details. If $F=0$, \eqref{eq:folded_chain_equation_motion_y} becomes
\begin{equation}\label{eq:folded_chain_equation_motion_F_0}
(1-f\sgn(\ydot))\ydot^2+(g-2\yddot)(\ell-y)=0.
\end{equation}
It follows from \eqref{eq:folded_chain_equation_motion_F_0} that $\sgn(\ydot)\equiv+1$ whenever $\ydot(0)\geqq0$. If, in addition, $f=1$ then $\yddot=\frac12g$, which means that the free end of the chain falls freely. However, if $f<1$ then $\yddot>\frac12g$ and the chain falls faster. Introducing the scaled variables defined in \eqref{eq:falling_chain_eta_xi}, we write \eqref{eq:folded_chain_equation_motion_F_0} as
\begin{equation}\label{eq:folded_chain_equation_motion_eta}
(1-f){\eta'}^2+2(1-\eta'')(1-\eta)=0,
\end{equation}
valid whenever $\eta'(0)\geqq0$. Equation \eqref{eq:folded_chain_equation_motion_eta} can be solved by quadratures. Setting $\eta_0:=\eta(0)\geqq0$ and letting $\eta'(0)=0$, we arrive at the following (implicit) representation for the solution,
\begin{equation}\label{eq:folded_chain_xi_of_eta}
\xi=\sqrt{1-\frac12f}\int_{\eta_0}^\eta\frac{1}{\sqrt{(1-\eta_0)^{2-f}(1-x)^{f-1}-1+x}}dx,
\end{equation}
and correspondingly $P$ in \eqref{eq:folded_chain_P} becomes
\begin{equation}\label{eq:folded_chain_P_eta}
\frac{P}{P_0}=\frac{1+f}{2-f}\left[(1-\eta_0)^{2-f}(1-\eta)^{f-1}-1+\eta\right]+\eta.
\end{equation}

To illustrate this solution we consider the case when $\eta_0=0$, so that the chain is fully extended upward before being set free. The time of fall $\tf$, which is delivered by the formula
\begin{equation}\label{eq:folded_chain_tf}
\frac{\tf}{t_0}=\sqrt{1-\frac12f}\int_0^1\frac{1}{\sqrt{(1-x)^{f-1}-1+x}}dx,
\end{equation}
is plotted in Fig.~\ref{fig:folded_chain}(b) as a function of $f$. Again $\tf$ is equal to the time of free fall $t_0$ only for $f=1$; otherwise, $\tf<t_0$. Figure~\ref{fig:folded_chain}(c) shows how $y$ grows in time for three values of $f$. Here, unless $f=1$, $\ydot$ diverges to $+\infty$ as $t\to\tf$. In Fig.~\ref{fig:folded_chain}(d) are plotted in a logarithm scale the graphs of $P$ as delivered by \eqref{eq:folded_chain_P_eta}. For $f=1$, $P=3P_0\eta$, so that the dynamic weight of the completely fallen chain is $\Pf=3P_0$. For $f<1$, $P$ diverges to $+\infty$ as $\eta\to1$ and each graph in Fig.~\ref{fig:folded_chain}(d) is plotted up to the time when $\eta=1-\varepsilon$, where $\varepsilon$ is the same cutoff parameter introduced in Section~\ref{sec:falling}.

\section{Sliding Chain}\label{sec:sliding}
\begin{figure}[!h]
  \centering
  \hspace{0.05\linewidth}
  \subfigure[]{\includegraphics[width=.3\linewidth]{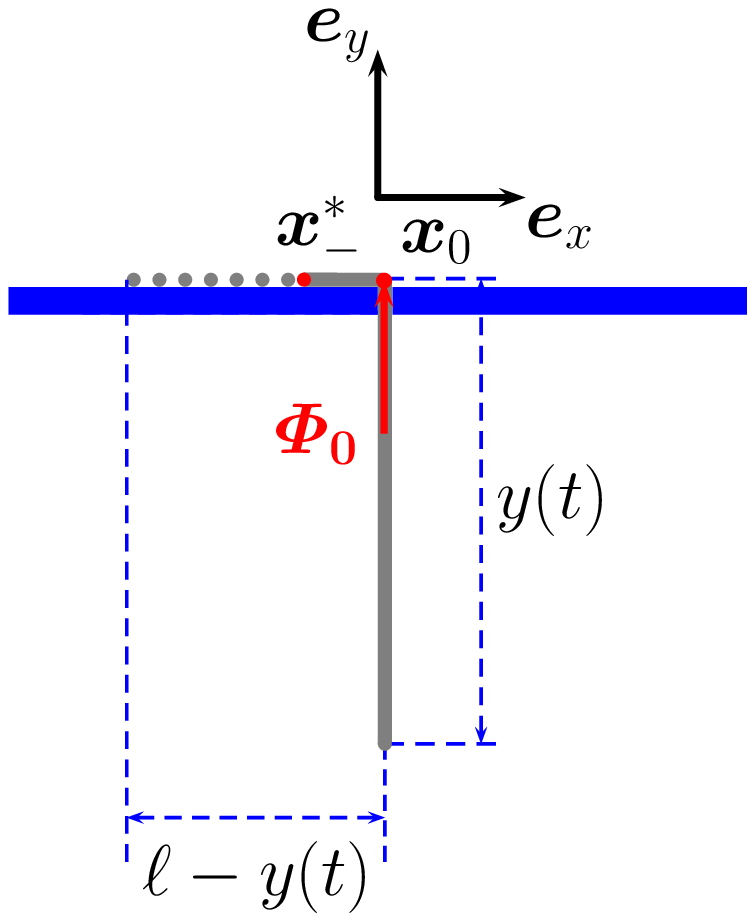}}
  \hspace{0.10\linewidth}
  \subfigure[]{\includegraphics[width=.5\linewidth]{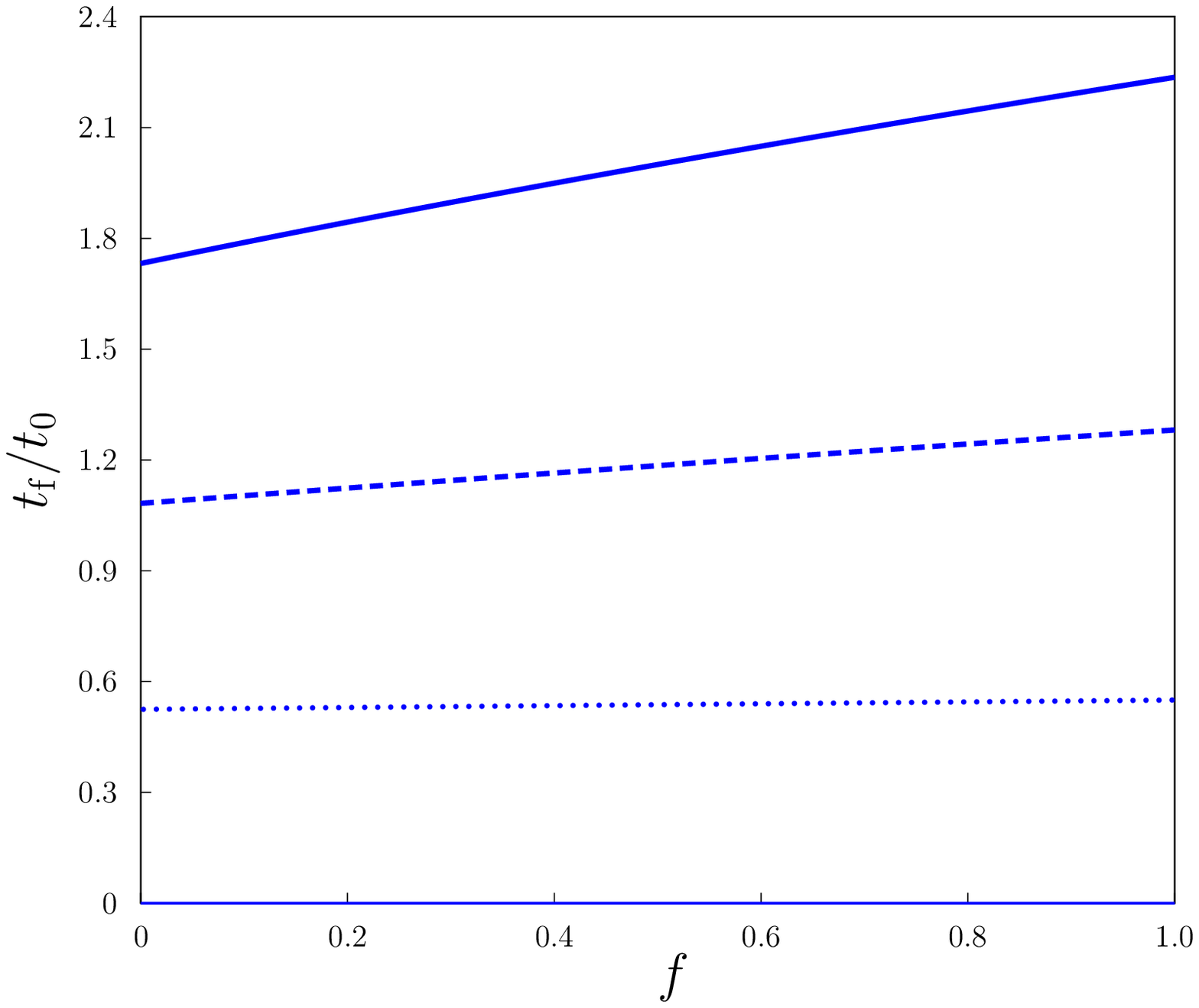}}
  \caption{(a) Idealized model for a sliding chain. Standing internal and external shocks are present at $\xnot$ and $\xstarm$, respectively. The length of the chain fallen at time $t$ is denoted by $y(t)$; $\ell$ is the total length of the chain. The smooth, rigid plane though which the chain slides opposed a vertical impulsive reaction $\bm{\Phi}_0=\Phi_0\ey$. (b) The time of fall $\tf$ (scaled to the time of free fall $t_0$ in \eqref{eq:falling_chain_t_0}) is plotted as a function of $f$ according to \eqref{eq:sliding_chain_xi_of_eta} for $\eta_0=0$ (solid line), $\eta_0=0.25$ (dashed line), and $\eta_0=0.75$ (dotted line). As $\eta_0\to1$, $\tf\to0$ for all $f$ (thin solid line).
}
\label{fig:sliding_chain}
\end{figure}

An idealized sliding chain is represented in in Fig.~\ref{fig:sliding_chain}(a).
Here a heap of links, whose specific shape and structure is ignored, is sitting on a smooth, rigid plane and act as a reservoir for the chain sliding away under gravity through a hole in the plane.\footnote{Were the chain on the plane taken to be in a definite shape at the start, say arranged along a straight line, our problem would change considerably. As shown in \cite{hanna:slack}, there is no guarantee that the chain, once set in motion, would keep sliding on the plane before plunging down into the hole. Our assumption here makes the problem akin to that envisaged by Cayley~\cite{cayley:class}.} An internal shock travels through the singular point $\xnot$ as does an external shock through $\xstarm$, which we take to be extremely close to $\xnot$. The origin of the arch-length co-ordinate $s$ is such that the falling free end of the chain has $s=\ell$, where $\ell$ is the total length of the chain. As above, we denote by $\sstarm(t)$ and $s_0(t)$ the propagating shocks' fronts. With this choice for the orientation of $s$, at $\xnot$ the unit tangent $\ut$ jumps from $\tm=\ex$ to $\tp=-\ey$.

Kinematic compatibility acquires again the form \eqref{eq:falling_chain_kinematic_compatibility}, both shocks are standing in space, and \eqref{eq:falling_chain_shock_speed} is replaced by
\begin{equation}\label{eq:sliding_chain_shock_speed}
\ydot=-\sspeed,
\end{equation}
where $y(t)$ denotes the length of the chain fallen at time $t$. Accordingly, \eqref{eq:jump_tension} becomes
\begin{equation}\label{eq:sliding_chain_jump_tension}
\jtau=\lambda f\sgn(\ydot)\ydot^2.
\end{equation}
Since the tension vanishes at the free end of the chain, integrating \eqref{eq:balance_equation_linear_momentum} we readily arrive at
\begin{equation}\label{eq:sliding_chain_tension_plus}
\taup=\lambda(g-\yddot)y.
\end{equation}
As shown in Fig.~\ref{fig:sliding_chain}(a), the plane exerts an impulsive reaction, $\bm{\Phi}_0=\Phi_0\ey$, with $\Phi_0\geqq0$. At $\xnot$ the balance equation for linear momentum  \eqref{eq:evolution_equation_v_0} then requires that
\begin{subequations}\label{eq:sliding_chain_tau_m_and_Phi_0}
\begin{align}
\taum &=\lambda\ydot^2,\label{eq:sliding_chain_tau_m}\\
\Phi_0 &=\taup-\lambda\ydot^2.\label{eq:sliding_chain_Phi_0}
\end{align}
\end{subequations}

Combining together \eqref{eq:sliding_chain_tau_m}, \eqref{eq:sliding_chain_tension_plus}, and \eqref{eq:sliding_chain_jump_tension}, we obtain the equation of motion for $y$,
\begin{equation}\label{eq:sliding_chain_equation_motion}
(1+f\sgn(\ydot)\ydot^2+(\yddot-g)y=0,
\end{equation}
and
\begin{equation}\label{eq:sliding_chain_Phi_0_explicit}
\Phi_0=\lambda f\sgn(\ydot)\ydot^2,
\end{equation}
which delivers an admissible reaction only is $\ydot\geqq0$, which is indeed a consequence of \eqref{eq:sliding_chain_equation_motion}, if $\ydot(0)\geqq0$ and $y(0)>0$, both taken as valid here. For $f=0$ equation \eqref{eq:sliding_chain_equation_motion} coincides with both Cayley's equation \cite{cayley:class} and the equation proposed in \cite[p.\,257]{sommerfeld:mechanics}. From \eqref{eq:sliding_chain_Phi_0_explicit}, we also derive the dynamical weight $P$ of the chain,
\begin{equation}\label{eq:sliding_chain_P}
P=\lambda f\ydot^2+\lambda(\ell-y)g.
\end{equation}
Identifying $\taum$ with the tension $\taustarm$ at $\xstarm$, by \eqref{eq:sliding_chain_tau_m} and \eqref{eq:external_shock_balance_linear_momentum_plus_minus}, we also conclude that $\Phistarm=\zero$, and so the heap of links suffers no horizontal force as a consequence of the chain's motion.

By \eqref{eq:external_shock_balance_energy_plus_minus} and \eqref{eq:sliding_chain_tau_m}, we easily see that the power $\Wstarm$ expended at the external shock is
\begin{equation}\label{eq:sliding_chain_W_star}
\Wstarm=-\frac12\lambda\ydot^3,
\end{equation}
while by \eqref{eq:W_s_formula_chain} the power expended at the internal shock is
\begin{equation}\label{eq:sliding_chain_W_s}
\Ws=-\lambda f\ydot^3.
\end{equation}
Since the total kinetic and potential energies are here expressed as
\begin{equation}\label{eq:sliding_chain_K_V}
K=\frac12\lambda y\ydot^2\quad\text{and}\quad V=-\frac12\lambda gy^2,
\end{equation}
respectively, a simple computation shows that
\begin{equation}\label{eq:sliding_chain_energy_total_balance}
\dot{K}+\dot{V}=\Wstarm+\Ws
\end{equation}
along every solution of \eqref{eq:sliding_chain_equation_motion}, as expected.

A special solution of \eqref{eq:sliding_chain_equation_motion} deserves notice by its simplicity. If $y(0)=0$ and $\ydot(0)=0$, then either $y\equiv0$ or $y=\frac12at^2$ with
\begin{equation}\label{eq:sliding_chain_a}
a=\frac{g}{3+2f},
\end{equation}
which corresponds to a uniformly accelerated fall with acceleration $\frac{g}{5}\leqq a\leqq\frac{g}{3}$. Correspondingly, the time $\tf$ for the complete fall of the chain is $\tf=t_0\sqrt{3+2f}$, where $t_0$ is the same as in \eqref{eq:falling_chain_t_0}, and
\begin{equation}\label{eq:sliding_chain_P_zero}
\frac{P}{P_0}=1-\frac{3}{3+2f}\left(\frac{t}{\tf}\right)^2,
\end{equation}
where $P_0$ is again the total static weight of the chain.

More generally, introducing the same scaled variables as in \eqref{eq:falling_chain_eta_xi}, we give \eqref{eq:sliding_chain_equation_motion} and \eqref{eq:sliding_chain_P} the following forms
\begin{equation}\label{eq:sliding_chain_equation_motion_eta}
(1+f){\eta'}^2+(\eta''-2)\eta=0,
\end{equation}
\begin{equation}\label{eq:sliding_chain_Phi_eta}
\frac{P}{P_0}=\frac12f{\eta'}^2+1-\eta,
\end{equation}
both valid for $\eta'\geqq0$. Letting $\eta_0:=\eta(0)>0$ and $\eta'(0)=0$, the implicit solution of \eqref{eq:sliding_chain_equation_motion_eta} reads as
\begin{equation}\label{eq:sliding_chain_xi_of_eta}
\xi=\frac12\sqrt{3+2f}\int_{\eta_0}^\eta\frac{x^{1+f}}{\sqrt{x^{3+2f}-\eta_0^{3+2f}}}dx,
\end{equation}
valid for $\eta_0\leqq\eta\leqq1$. The corresponding time of fall $\tf$ is plotted in Fig.~\ref{fig:sliding_chain}(b) for different values of $\eta_0$.


\section*{Acknowledgments}
I am indebted to Peter Palffy-Muhoray for having sparkled my recent interest in chain dynamics and to James A. Hanna for having encouraged me to pursue further the method of \emph{external shocks} presented in this paper (of course, all possible mistakes remain my own.)



\end{document}